\documentclass[useAMS,usenatbib]{mnras}

\usepackage[T1]{fontenc}

\usepackage{graphicx}	
\usepackage{amsmath}	
\usepackage{amssymb}	
\usepackage{verbatim} 
\usepackage{lscape}
\usepackage{dcolumn}
\usepackage{color}
\usepackage{float}
\usepackage{subfig}
\usepackage{tabularx}
\usepackage{soul}
\usepackage{ulem}
\definecolor{jm}{rgb}{0.23, 0.53, 0.75}

\definecolor{np}{RGB}{0, 255, 0}
\definecolor{js}{RGB}{0, 0, 0}
\definecolor{red}{RGB}{255, 0, 0}
\newcommand{\JS}[1]{\textcolor{js}{#1}} 
\newcommand{\NP}[1]{\textcolor{js}{#1}}
\newcommand{\JSH}[1]{\textcolor{js}{#1}}

\title[Extended primordial black hole mass functions]{Press-Schechter primordial black hole mass functions and their observational constraints}

\author[Sureda et al.]{
Joaqu\'in Sureda,$^{1,2}$\thanks{E-mail: jmsureda@uc.cl}
Juan Maga\~na,$^{1,2}$
Ignacio J. Araya$^{3}$
and Nelson D. Padilla$^{1,2}$
\\
$^{1}$Instituto de Astrof\'\i sica, Pontificia Universidad Cat\'olica de Chile, Vicu\~na Mackenna 4860, Santiago, Chile\\
$^{2}$Centro de Astro-Ingenier\'\i a, Pontificia Universidad Cat\'olica de Chile, Vicu\~na Mackenna 4860, Santiago, Chile\\
$^{3}$Instituto de Ciencias Exactas y Naturales, Facultad de Ciencias, Universidad Arturo Prat,\\ Avenida Arturo Prat Chac\'on 2120, 1110939, Iquique, Chile
}

\date{Accepted XXX. Received YYY; in original form ZZZ}

\pubyear{2015}

\begin{document}
\label{firstpage}
\pagerange{\pageref{firstpage}--\pageref{lastpage}}
\maketitle

\begin{abstract}

We present a modification of the Press-Schechter (PS) formalism to derive general mass functions for primordial black holes (PBHs), considering their formation as being associated to the amplitude of linear energy density fluctuations. To accommodate a wide range of physical relations between the linear and non-linear conditions for collapse, we introduce an additional parameter to the PS mechanism, and that the collapse occurs at either a given cosmic time, or as fluctuations enter the horizon. We study the case where fluctuations obey Gaussian statistics and follow a primordial power spectrum of broken power-law form with a blue spectral index for small scales. 
We use the observed abundance of super-massive black holes (SMBH) to constrain the extended mass functions taking into account dynamical friction.  We further constrain the modified PS by developing a method for converting existing constraints on the PBH mass fraction, derived assuming monochromatic mass distributions for PBHs, into constraints applicable for extended PBH mass functions.
We find that when considering well established monochromatic constraints there are regions in parameter space where all the dark matter can be made of PBHs. Of special interest is the region for the characteristic mass of the distribution $\sim 10^2M_\odot$, for a wide range of blue spectral indices in the scenario where PBHs form as they enter the horizon, where the linear threshold for collapse is of the order of the typical overdensities, as this is close to the black hole masses detected by LIGO which are difficult to explain by stellar collapse.
\end{abstract}




\section{Introduction}

In the last decades, the concordance cosmological model that includes a cosmological constant and cold dark matter ($\Lambda$CDM) has been established 
to explain the growth of large-scale structures and the late 
accelerating expansion of the Universe. Under this paradigm, the dark matter (DM) is cold and made up of non-relativistic and collisionless particles which behave as a pressureless
fluid \citep{Planck:2018}. There are several candidates for the cold dark matter component, including weakly interacting massive particles \citep[WIMP's,][]{Arcadi:2018EPJC,wimps:2019}, axions \citep{Peccei:1977,Marsh:2016}, or ultra-light axions \citep{Hu:2000PhRvL, Schive:2014} which can be described by a coherent scalar field \citep{Matos:2000ss, Matos:2009}, among others \citep{Feng:2010}. Nevertheless, there are no direct astrophysical observations or accelerator detections of these particles and the nature of dark matter is still unknown \citep{Liu:2017}. 

An alternative hypothesis is to consider that primordial black holes (PBHs) are an important fraction (or all) of DM \citep[see][for recent reviews]{Khlopov_2010,Carr:2020arXiv200212778C,Carr:2020b,Green:2020}. In pioneer works, \citet{Zeldovich1966} and \citet{Hawking1971}  \citep[see also][]{Carr_Hawking:1974} discussed the possibility that overdensities in early stages of the Universe could collapse forming PBHs (see also \citealt{Carr1975} for PBH collapsing from cosmic strings). The idea of PBHs as the nature of dark matter has regained interest
with the recent detection by the Laser Interferometer Gravitational-Wave Observatory \citep[LIGO,][]{Abbott:2016blz} of gravitational waves (GW) produced by the merger of a pair of black holes with masses $\sim 30 M_{\odot}$ whose origin could be primordial \citep{Bird2016, Garcia-Bellido:2017fdg, Sasaki:2018,Jedamzik:2020}. Although the standard scenario for the PBH formation is the collapse of overdense fluctuations which exceed a threshold value when they re-enter to the horizon, other mechanisms involving phase transitions or topological defects have been proposed for the PBH production in the inflationary/radiation epochs, for instance, collapse of cosmic strings \citep{Hogan:1984PhLB,Hawking:1989PhLB,Polnarev:1991,Nagasawa:2005hv}, collapse of domain walls \citep{Rubin:2000dq, Rubin:2001yw,Liu:2019lul,Ge:2019ihf}, bubble collisions \citep{Hawking:1982_bubble,Kodama:1982sf,Deng:2017uwc, Deng:2020arXiv}, and softening of the equation of state \citep{Canuto:1978MNRAS,Khlopov:1980PhLB}, among others.

PBHs can have any range of masses since they are not restricted to form from dying stars. However, the minimum possible mass of a BH can be estimated considering that for any lump of mass $m$, in order to form a black hole, its Compton wavelength, $\lambda_{C}=h/mc$, has to be smaller than its Schwarzschild radius. This leads to the lower bound of one Planck mass $M_{PBH}\sim 10^{-5}$g. The upper mass could in principle be as large as $\sim 10^{50} g$ in some PBH formation scenarios \citep[see][and references therein]{Carr:2020arXiv200212778C, Carr:2020b}. 

Given that low mass PBHs can be close to their last evaporation stages via Hawking radiation \citep{Hawking:1974Nature,Hawking:1975}, this introduces interesting prospects for their detection \citep{Laha:2019PhRvL, Ballesteros:2019}. For instance it is possible that the evaporation radiation affects the HI content of the universe at redshifts prior to the formation of the first stars (e.g. \citealt{Mack2008}).
As PBHs are at least several orders of magnitude more massive than the most massive particles, they would constitute an extremely cold type of dark matter. As pointed out by \citet[][see also \citealt{Angulo:2017}]{Angulo2010}, the early epochs of decoupling of neutralinos (candidates for DM) from radiation makes for the possibility of dark matter haloes with masses as low as one Earth mass.  PBHs can therefore form haloes of even lower masses \citep{Tada:2019, Niikura:2019,Scholtz:2019,AxionPBH:2020arXiv200107476H,Hertzberg:2020}.  These haloes of PBHs would emit radiation as their smaller members evaporate, and this could in principle be detected with current and future high energy observatories such as Fermi and the Cherenkov Telescope Array (e.g. \citealt{Fermi-LAT:2018pfs, CTA:2013APh})

Estimates of the fraction of  dark matter in PBHs ($f$) at different mass windows can be obtained  from evaporation by Hawking radiation \citep{Hawking:1974Nature,Hawking:1975} and from their gravitational/dynamical effects, including GW observations \citep[see][]{Carr:1999ApJ_dynamical,Carr:2016,Wang:2018,Carr:2020arXiv200212778C,Carr:2020b}. Some of the stronger constraints for evaporating PBHs are imposed by standard big bang nucleosynthesis (BBN) processes and the extragalactic $\gamma$-ray background radiation \citep{Carr:2016_gammaray, Keith:2020}.  Other bounds on $f$ are obtained from the gravitational lensing effects of background sources (for instance stars in the Magellanic clouds) due to PBHs \citep{Green:2016PhRvD,Niikura:2019, Inoue:2018}. \citet{Hawkins:2020arXiv200107633H} found a low probability for the observed microlensing of QSOs by stars in lensing galaxies and argued that an intriguing possibility is the lensing by PBHs. Another limit is provided by the capture of PBHs by stars, white dwarfs or neutron stars \citep{Capela:2013}. Recently, \citet{Scholtz:2019} explore the capture probability of a PBH with $\sim 10$ earth masses by the Solar system as an alternative for the hypothetical planet nine. On the other hand, a passing PBH or PBH clumps could disrupt globular clusters and galaxies in clusters \citep{Carr:1999ApJ_dynamical,Green:2016PhRvD, Carr:2020b}. Although
there are several observational constraints on $f$,
most of them are for monochromatic PBH mass distributions.  These can be turned into constrains for extended PBH mass distributions following certain statistical procedures  \citep[see for instance][]{Carr:2017, Bellomo_2018}

 One of the simplest approaches to determine the PBH mass distribution assumes that there is a characteristic scale, $\lambda_{c}$, for the fluctuations which collapse to a PBH; i.e. the density fluctuations have a monochromatic power spectrum, and hence all the  PBHs  have $M\sim M_{c}$ i.e. a monochromatic mass function. A natural extension is to consider that PBHs form in a wider range of masses (for instance from particular inflationary scalar field potentials), which was pursued by several authors  \citep{Dolgov:1993PhRvD, Garcia-Bellido:1996PhRvD, Clesse:2015PhRvD, Green:2016PhRvD, Inomata:2017PhRvD,Inomata:2018, Luca:2020} and a with steep power-law power spectrum for density fluctuations \citep{Carr1975}. 
 
 Early works (e.g. \citealt{Peebles:1970ApJ}) showed that the index $n$ of the primordial power spectrum $P(k)\propto k^n$ should be close to $n\sim 1$ in order for there to be homogeneity on large scales.
This primordial power spectrum is referred to as the scale-invariant Harrison-Zeldovich-Peebles spectrum \citep{Harrison:1970,Zeldovich:1972spec,Peebles:1970ApJ} and received further support when inflation was proposed as a possible  stage of the very early Universe \citep{Guth1981}.  However, this prevents PBHs formed by direct collapse to constitute a sizeable fraction of the dark matter in the Universe \citep{Carr1975, Josan:2009, Green:1999PhRvD}.

Therefore, to increase the abundance of PBHs it is necessary to enhance the amplitude of the primordial power spectrum on specific scales.  For instance, hybrid inflation models can provide spectral indices greater than one, i.e. blue spectral indices \citep{Linde:1994PhRvD}. \citet{Kawasaki:2013PhRvD} studied PBH formation and abundance in an axion-like curvaton model with a blue-tilt ($n_{b}\sim 2-4$) in the power spectrum of primordial curvature perturbations \citep[see also][]{Gupta:2018}.

The probability distribution of density fluctuations in the early universe, typically assumed to be  Gaussian, can also play an important role in the PBH production. Several authors have looked into the effect of non-Gaussian distributions, showing that these introduce changes in the abundance (enhancement or suppression) and clustering of PBHs, and hence they also change their allowed fraction as an energy component of the Universe \citep{Bullock:1997PhRvD, Hidalgo:2007, Young:2013JCAP, Young:2015JCAP,Franciolini2018}. 

In this paper, we investigate PBH formation in a modified Press-Schechter \JS{\citep[PS from now on,][]{PS:1974ApJ}} formalism  \JS{that relates the amplitude of linear energy density perturbations to PBH abundance. PS has been widely used to estimate the mass distributions of gravitationally collapsed dark matter haloes, where the relation between the linear overdensity and the physical collapse are known via the spherical collapse model and its subsequent improvements. As this model is not applicable to PBHs, additional parameters and considerations are needed.} \JS{Our first new parameter in the PS formalism for PBHs is the fraction of the linear fluctuation to undergo collapse.  This allows to accommodate the widest variety of physical connections between the linear fluctuations and the actual physical condition for PBH formation. We also relate PBH formation with linear perturbations at a single epoch or at the time the fluctuation scale re-enters the horizon. 
We show how the modified PS formalism leads to extended PBH mass functions starting from a primordial power spectrum (PPS) of fluctuations with a broken power-law form with enhanced power on small scales. }

In Section \ref{sec:broad_mass_function_formalism}, we use a modified PS formalism to derive an extended PBH mass function starting from a broken power-law primordial power spectrum \JS{for two different PBH formation timings}. The functional form of the obtained PBH mass function is described by a type of Schechter function with a power-law slope and an exponential cutoff. In Section \ref{sec:constraints}, we introduce a new constraint for extended mass distributions looking at super massive black holes and also a new statistical analysis method to turn existing constraints on the PBH mass fraction $f$ coming from monochromatic distributions into constraints for extended PBH mass functions. We then consider a series of monochromatic constraints and use them to derive the corresponding ones for the PS mass functions obtained in this work. In Section \ref{sec:results}, we show the resulting fractions $f$ for different choices of the Schechter function parameters, and we show that there are regions in parameter space where the entirety of the DM can be made up by PBHs. Finally, in Section \ref{sec:conclusions}, we summarise our main results and conclusions. 
Throughout this work we assume a flat cosmology with $\Omega_{m,0} = 0.315$  ;  $\Omega_{dm,0} = 0.264$  ;  $\Omega_{r,0} = 9.237\times 10^{-5}$  and a Hubble constant $H_{0}=67.36\, \mathrm{km \,s^{-1} Mpc^{-1}}$ consistent with measurements from the Planck satellite \citep{Planck:2018}.

\section{The Primordial Black Hole Mass Function}\label{sec:broad_mass_function_formalism}

PBH formation is related to the density fluctuations in the early universe, which are quantified by the Primordial Power Spectrum. The standard PPS is parametrised by a power-law
\begin{equation}
P(k)=A_s \left(\frac{k}{k_0}\right)^{n_s},
\label{eq:pkplaw}
\end{equation}
\noindent
where $A_s$ is an arbitrary normalisation and $n_{s}$ is the spectral index. These parameters
are measured by the Planck collaboration at $k_0=0.05\,\mathrm{ Mpc}^{-1}$, obtaining $A_s = 2.101\times 10^{-9} \,\mathrm{ Mpc}^3$ 
and $n_{s}=0.9649 \pm 0.0042$, i.e. a red-tilt power spectrum with no evidence for significant deviation of the power-law at $0.008 \mathrm{Mpc^{-1}} \lesssim k \lesssim 0.1\mathrm{Mpc^{-1}}$ \citep{Planck:2018}.

We also consider a different PPS with a blue tilted spectrum at small scales, which will be referred to as the broken primordial power spectrum (BPPS). It is defined as

\begin{equation}
P(k) = \begin{cases}
A_s\left(\frac{k}{k_0}\right)^{n_s} &\text{for $k<k_{piv}$},\\
A_s\epsilon\, \left(\frac{k}{k_0}\right)^{n_{b}} &\text{for $k\geq k_{piv}$},
\end{cases}
\label{eq:pkbroken}
\end{equation}

\noindent
where $k_{piv}$ is the pivot wavenumber above which the spectral index is blue, i.e. $n_{b}>1$ and $\epsilon$ is a constant introduced to ensure the continuity of $P(k)$, defined as

\begin{equation}
    \epsilon=\left(\frac{k_{piv}}{k_0}\right)^{n_s-n_b}.
\end{equation}

\begin{figure}
  \centering
  \includegraphics[width=0.45\textwidth]{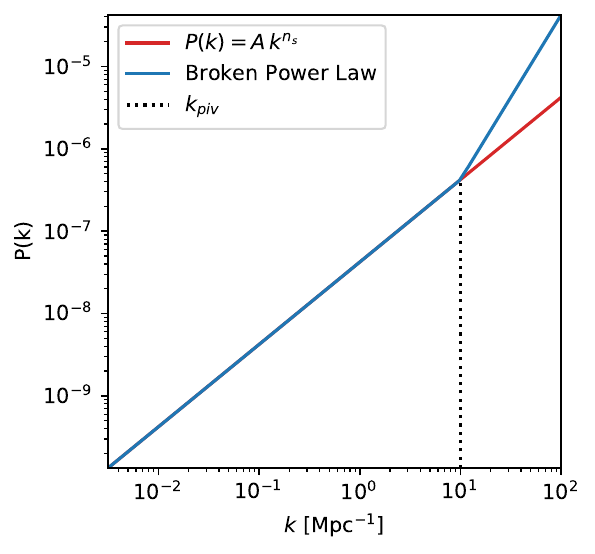}
\caption{Primordial power spectrum, $P(k)$ as function of the wavenumber $k$ for two cases, the power law given by Eq. \eqref{eq:pkplaw} and the broken power law given by Eq. \eqref{eq:pkbroken}. We have considered $n_{s}=0.9649$ for the power law spectrum and $n_{b}=2$ with $k_{piv}=10\,\mathrm{Mpc}^{-1}$ for the broken power law.} 
\label{fig:powerspectrum}
\end{figure}

Figure \ref{fig:powerspectrum} shows the power spectrum as function of the wavenumber for a power law (Eq. \eqref{eq:pkplaw}) and the broken power-law (Eq. \eqref{eq:pkbroken}). Notice that for the latter, there is an enhancement for wavenumbers above than the pivot scale. Throughout this work, we have set $k_{piv}=10\,\mathrm{Mpc}^{-1}$. Notice that
our choice of $k_{piv}$ is conservative \citep[see for example][]{Hirano:2015wla}. Future CMB experiments as Primordial Inflation eXplorer (PIXIE) will be able to constrain the primordial power spectrum at wavenumbers between $50 \text{Mpc}^{-1}\lesssim k \lesssim 10^{4} \text{Mpc}^{-1}$ \citep{Chluba:2012MNRAS}.


To obtain the mass function of primordial black holes, we adopt the formalism by \cite{SMT2001}, which is a Press-Schechter approach that solves the peaks-within-peaks underestimate of the abundance of objects (another approach to calculate the PBH abundance is the peaks theory, see for instance \citealt{Bardeen:1986,Green:2004PhRvD,Young:2014J,Germani:2019,YoungandMusso:2020}). In this approach, the abundance depends on the linear overdensity above which objects collapse and form.  The use of linear overdensities is what allows to use the PS formalism in the first place, as it adopts Gaussian statistics. \JS{There are similarities and differences with the formalism followed for dark matter haloes \citep[briefly discussed by][]{YoungandMusso:2020}. We start highlighting the similar aspects.} 

Following \JS{the standard} PS formalism, the extended PBH mass function is defined as

\begin{equation}
\left(\frac{dn}{dM}\left( M\right) \right)_{PS}  =\nu f\left(  \nu\right)  \frac{\rho_{DM}}{M^{2}}\frac{d\log\nu}{d\log M}=f\left(  \nu\right)  \frac
{\rho_{DM}}{M}\frac{d\nu}{dM},
\label{eq:dndm_gen}
\end{equation}
\noindent
where $\rho_{DM}$ is the dark matter density, and $\nu(M)$ can be interpreted as the peak height defined as

\begin{equation}
\nu\left(  M\right)
=\frac{\delta_{c}}{\sigma\left(  M\right) },
\label{eq:nu}
\end{equation}
being $\delta_{c}$ \JS{the linear} threshold density contrast \JS{(or critical density contrast)} for PBH formation and $\sigma(M)$ the variance of the density field. Notice that $\delta_{c}$ is the linear overdensity and its relation with the non-linear density (the physical one) depends on the physical mechanism by which a region collapses into a PBH. The relation between the linear and non-linear overdensity has been investigated by several authors \citep[see for instance][]{Yoo:2018,Kawasaki:2019PhRvD, Young:2019JCAP, Luca:2019JCAP, Kalaja:2019JCAP,Musco:2019} \JSH{. For instance, \citet{Musco:2021} give a general prescription to obtain the non-linear critical density contrast. However,} its calculation is beyond the scope of this work. \JSH{Additionally,} for the linear perturbations to be related to the physics of collapse, $\delta_c$ should be ideally similar, or of the order, of the typical amplitude of fluctuations $\left<|\delta|^2\right>^{1/2}$ at the time of collapse. \JSH{Note that even if $\delta_c \ll \left<|\delta|^2\right>^{1/2}$, there is still a way to relate the perturbations with the shape of the PPS.} We will check whether there are windows in the extended PS parameter space that allow PBHs as dark matter where this condition is met.

On the other hand, \JS{in PS it is assumed} that the distribution of $\sigma$
follows Gaussian statistics, as
\begin{equation}
f(\nu(M))=\frac{2}{\sqrt{2\pi}}\exp \left(-\frac{1}{2}  \nu(M)^{2} \right).
\label{eq: fnu Gaussian}
\end{equation}

The variance of the fluctuations
of a smoothed density field is given by
\begin{equation}
\sigma^{2}\left(  M\right)  =4\pi\, D^2(a)%
{\displaystyle\int\limits_{0}^{\infty}}
k^{2}P\left(  k\right)  \widehat{W}^{2}\left(  R\left(  M\right)  ,k\right) dk,
\label{eq:sigmasquared}
\end{equation}
where $k$ is the comoving wavenumber, $P(k)$ is the primordial power spectrum, $ D(a)$ is the growth factor of the fluctuations  at certain scale factor $a$, and $\widehat{W}$ corresponds to a window function. Several authors have investigated the effect of the choice of $\widehat{W}$ on different power spectra and hence on the \JS{halo or} PBH abundance under the PS and peaks theory approaches \citep{Gow:2020a,Young:2020IJMPD}. For instance, \citet{Gow:2020a} found that, by considering a top-hat or a Gaussian smoothing function in a log-normal power spectrum, the amplitude difference in a range of masses is $\lesssim 20\%$ and the resulting mass distributions are very similar. 
For simplicity, we use the sharp (top-hat) $k$-space window function 

\begin{equation}
\widehat{W}^{2}\left(  R\left(  M\right)  ,k\right)  =\left\{
\begin{array}
[c]{cc}%
1, & k\leq k_{R}\\
0, & k>k_{R}%
\end{array}
\right.  ,
\end{equation}
\noindent 
where 
\JS{
\begin{equation}
    k_R = \frac{2\pi}{R(M)},
    \label{eq: k_R(M) definition}
\end{equation}}
$R$ is the comoving radius, and its dependence with the mass $M$ is given by
\begin{equation}
    M(a) = \frac{4\pi}{3}(a\,R)^3 \bar{\rho}(a),
    \label{eq: Horizon Mass}
\end{equation}
where $\bar{\rho}$ is the background density.
The mass defined in Eq. \eqref{eq: Horizon Mass} corresponds to the total energy density within a sphere of radius $R$ as a function of the scale factor $a$. 

\JS{Notice that} here we find the first difference between PS applied to PBH formation compared to halo formation. We want to associate a certain scale $k_R$ with the mass of a PBH rather than with the total energy density. This can be interpreted in two ways\JS{, i)} only a small fraction of the energy density within the horizon will undergo collapse\JS{; ii)} only a fraction of the modes with wavenumber $k$ will collapse forming a PBH. \NP{One can in principle consider any combination of these two scenarios bearing in mind that from a statistical point of view option (i) retains the strongest connection between the linear fluctuations and the actual non-linear collapse. 
These two possibilities can fit into the equations by proposing,}
\begin{equation}
    M_{PBH} = f_m \times M_H = \frac{4\pi}{3}(a\,R)^3 \,f_m\,\bar{\rho}(a),
\end{equation}
\noindent
where $f_m$ is the fraction of the energy density that collapses into a PBH and $M_H$ is given by Eq. \eqref{eq: Horizon Mass}. \JS{ Option (i) will result from considering $f_m = \beta$, where $\beta$ corresponds to the fraction of energy density in the form of PBHs at the formation time \citep{Carr:2010},
\begin{equation}
    \beta(a) = \frac{\rho_{PBH}(a)}{\rho(a)};
    \label{eq:beta_def}
\end{equation}}
while option (ii) will occur for $f_m = 1$, implying that $f_m \in [\beta,1]$. Notice that the value of $f_m$ is degenerate with the energy density and, therefore, with the scale factor $a$.

Notice that $f_m$ is a different quantity than the fraction of the non-linear overdensity that undergoes collapse \citep[$\gamma$ in][for example]{Musco:2019}. While the latter is related to the physical collapse in the non-linear regime, $f_m$ arises from the assumption that there is an unknown relation between a non-linear overdensity and the linear overdensity. Additionally, we assume that $f_m$ is spatially and temporally constant. As $f_m$ can take values from $\beta$ to one, it does not determine the relationship between the energy contained in the physical overdensity and that contained in the PBH which is formed from its collapse. In particular, there are two factors, the fraction of the mass that collapses inside a linear fluctuation, and the fraction of linear fluctuations that result in a collapse. $f_m$ is only the former. This allows us to leave $f_m$ as a free parameter to investigate if it is possible to have a significant fraction of DM as PBHs. In future studies one  can corroborate whether the choice of $f_m$ makes physical sense in terms of the actual, non-linear PBH collapse.

Note that  for haloes the background density from where they collapse evolves as the haloes themselves. In contrast, for the PBH formation, the background density $\bar{\rho}$ evolves as radiation, whereas the PBH density evolves as matter. 

Therefore, in contrast to the PS formalism for haloes that requires a single linear overdensity for collapse as its only parameter, the general PS for PBHs requires two parameters, $\delta_c$ and $f_m$, to encompass different possible relations between the linear overdensity and the actual physical overdensity for collapse. 

Having a PBH mass function, we can compute the PBH number density and mass density. 
The number density of PBHs with mass between $M_{min} < M < M_{max}$ is given by

\begin{equation}
n_{PBH}  =%
{\displaystyle\int\limits_{M_{min}}^{M_{max}}}
\frac{dn}{dM}(M')\, \mathrm{dM'}.
\label{eq:NPBH}
\end{equation}

Similarly, the PBH mass density, is defined as  

\begin{equation}
\rho_{PBH}  =%
{\displaystyle\int\limits_{M_{min}}^{M_{max}}} M'\,\frac{dn}{dM}(M')\,\mathrm{dM'}.
\label{eq:rho_pbh}
\end{equation}
\noindent

The average mass for the distribution can be computed as

\begin{equation}
\left\langle M\right\rangle_{PBH} =\frac{\rho_{PBH}}{n_{PBH}}=\frac{%
{\displaystyle\int\limits_{M_{min}}^{M_{max}}}
M'\frac{dn}{dM}(M')dM'}{%
{\displaystyle\int\limits_{M_{min}}^{M_{max}}}
\frac{dn}{dM}(M')dM'}.
\label{eq:average_mass}
\end{equation}

\JS{As we will be interested in determining which range of $\delta_c$ and $f_m$ allow a large fraction of DM in PBHs,} it is useful to discuss  the mass limits of the integrals \NP{since these determine the overall normalisation of the PBH mass functions}.  The lower mass limit $M_{min}$ is related to the emission via Hawking radiation \citep{Hawking:1974Nature,Hawking:1975} of a black hole. The evaporation lifetime, $\tau_{ev}$, for a black hole with mass $M$ is 
\begin{equation}
\tau_{ev}=\frac{5120 \pi G^{2}M^{3}}{\hbar c^{4}} \sim 10^{64} \left(\frac{M}{M \odot}\right)^{3} yr.
\label{eq: evaporation lifetime}
\end{equation} 

\noindent
The mass of a PBH in the last stages of evaporation depends mostly on the redshift of evaporation and we will refer to this mass as $M_{ev}(z)$. We typically set the minimum mass at $M_{min}=M_{ev}(z)$. \JS{In the treatment of the minimum mass we are assuming that BH evaporation is an instantaneous process. This is justified given that half of the PBH mass is lost only in the last eighth of its lifetime (see Appendix \ref{sec:AppendixBHevap}).}

Regarding the upper mass limit, since PBHs of the highest masses tend to be  rare, it is possible that they will not be found within the causal volume at early times. We quantify this by defining the cumulative number density of PBHs above any mass $M$ as

\begin{equation}
    n_{PBH}(>M)= \int_M^\infty \frac{dn}{dM'}dM',
    \label{eq:cumulative number density}
\end{equation}

\noindent
and the cumulative number of PBHs is given by 

\begin{equation}
    N(>M) = Vn_{PBH}(>M) = V\int_M^\infty \frac{dn}{dM'}dM',
    \label{eq:cumulative number}
\end{equation}

\noindent
where $V$ is the volume in a spherical region. We will refer to the mass at which the cumulative number density equals one PBH per horizon volume as $M_{1pH}(z)$,

\begin{equation}
    n(>M_{1pH}(z))=1/V_{Hubble}(z).
\end{equation}

\noindent
Notice that $M_{1pH}(z)$ is a function of redshift because the Hubble volume $V_{Hubble}$ increases with cosmic time, and it will enter in the calculations as the upper mass of PBHs that are in causal contact at redshift $z$, as is the case of fluctuations that enter the horizon. In our calculations we adopt $M_{max}=M_{1pH}(z)$. \JS{This upper limit will be important mostly in very high redshift calculations such as the epoch of nucleosynthesis, when the horizon size is much smaller than today which makes the minimum detectable comoving abundance of PBHs, $1/V_{Hubble}$, much higher.}

Due to the definition of the mass function in Eq. \eqref{eq:dndm_gen} 

\begin{equation}
    \rho_{DM} = \int\limits_{0}^{\infty} M'\,\left(\frac{dn}{dM}(M')\right)_{PS}\,dM',
\end{equation}
\noindent
but this includes PBHs that have already evaporated and PBHs that are not likely to be found within the causal volume. Therefore, we normalise the mass function by

\begin{equation}
A_{n} \ =\ \frac{\int ^{\infty }_{0} M\ \left(\frac{dn}{dM}\right)_{PS} \ dM}{\int ^{M_{1pH}( z=0)}_{M_{ev}( z=0)} M\ \left(\frac{dn}{dM}\right)_{PS} \ dM},
\label{eq:norm}
\end{equation}
\noindent
enforcing that the PBH mass function satisfies

\begin{equation}
    \rho_{DM} = \int_{M_{ev}(z=0)}^{M_{1pH}(z=0)} M'\,A_n\,\left(\frac{dn}{dM}(M')\right)_{PS}\,dM'.
    \label{eq:rho_normalized}
\end{equation}

From now on, when we use the mass function of PBHs, we are considering the normalised mass function that satisfies \eqref{eq:rho_normalized}.\footnote{We are aware that the computation of the normalisation depends on $M_{1pH}$, which in turn depends on $A_n$, meaning that this is an iterative process. However, this process converges after a few iterations.} This is

\begin{equation}
    \frac{dn}{dM}(M) = A_n\, \left(\frac{dn}{dM}(M)\right)_{PS},
    \label{eq: normalized mass function}
\end{equation}
\noindent
with the PS mass function as defined in \eqref{eq:dndm_gen}. \JSH{We emphasise that this normalisation factor does not take into account any evolution of the PBH distribution besides the evaporation (via Hawking Radiation) and the increasing maximum mass within the Horizon $M_{1pH}$. Some effects such as clustering of PBHs at early times can be important for monochromatic mass functions as studied in \citet[][]{Inman:2019}. However, for extended mass functions these effects have not been studied in detail and are beyond the scope of this work.}



An important point in the construction of the PBH mass function, is that the actual amplitude of linear fluctuations that are informative of the physical collapse of PBHs can, in principle, be taken either at a single epoch, or at different moments during the radiation domination era. In the first scenario, which we refer to as \textit{Fixed Conformal Time} (FCT), all PBHs form with masses that correspond to linear overdensities taken roughly the same epoch. Adopting a  scale factor of PBH formation allows us to make quantifications of the PBH mass function in this scenario. We adopt $a_{fct}\approx2.04\times 10^{-26} $, right after inflation,  \NP{unless otherwise stated}. Phase transitions \citep{Kolb:1990,Gleiser:1998, Rubin:2001yw,Jedamzik:1999PhRvD, Ferrer:2019PhRvL} could naturally provide such a mechanism as they are triggered by a change in the global conditions of the Universe. For more details or examples of this kind of scenarios, see \citet{Hawking:1982_bubble,Moss:1994,Khlopov:1998,Deng:2017uwc,Lewicki:2019, Deng:2020,Kusenko:2020,Deng:2020arXiv}, \JSH{where models like vacuum bubble nucleation or collisions are discussed, which can be considered as FCT-like scenarios}. The second scenario, called \textit{Horizon Crossing} (HC), consists of linking the formation of PBHs to the linear amplitude of fluctuations as the scale associated to the mass of PBHs enters the horizon. The main feature of this scenario is that smaller PBHs are formed first in the Universe and the more massive ones are formed later. This kind of scenario is well motivated \citep[see for example][\JSH{where they use different approaches, such as critical collapse, to derive a PBH mass function using peaks theory or Press Schechter, for example, even in non-Gaussian regimes.}]{Green:1997PhRvD,Green:1999PhRvD,Green:2004PhRvD,Green:2016PhRvD, Young:2013JCAP, Young:2015JCAP}, and is expected of any underlying PBH formation mechanism that requires the collapsing scale to be in causal contact. \JSH{Then, these kind of models can be considered as HC-like.}

The details of the physics of each formation mechanism are beyond the scope of this work. \JSH{To first order we consider the choice of FCT or HC affecting only the effective PPS, more accurately, we only change the effective value of $n_b$ and, consequently, the slope of the mass function. This will change other properties of the PBH distribution in turn, such as the formation scale factor which will depend strictly on the model used to describe the PBH formation.} We do notice that these two scenarios are extremes of a continuous range of possibilities, which one could in principle parametrise. However, to simplify the algebra we only look at each of these two extremes in detail. \JSH{Throughout the rest of the text we will refer to HC (FCT) as a HC-like (FCT-like) scenario.}

Summarising, different formation mechanisms and considerations for the PPS will give different results for the Press-Schechter PBH mass function. In the following sub-sections, we discuss how to obtain the value of a typical overdensity as a function of the scale factor, $\left<|\delta|^2\right>^{1/2}(a)$, which we will compare with $\delta_c$. Later we analyse the choice of $f_m=\beta$ and relate it with the calculations of $\left<|\delta|^2\right>^{1/2}(a)$. Finally, we obtain the mass function for the FCT and HC scenarios, considering a standard PPS and a broken PPS.

\subsection{Typical overdensity $\left<|\delta|^2\right>^{1/2}$}
\label{sec:app_ratiodeltas}

\JS{It should be noted that the value of $\delta_c$ alone does not give enough information regarding the collapse of an overdensity. In order for it to be meaningful, it should be compared to the amplitude of a typical density contrast $\left<|\delta|^2\right>^{1/2}$ at the epoch or scale of interest.}


\JS{In the FCT scenario where the linear fluctuations are analysed at a fixed scale factor, the  scale of interest is the one associated to the mean mass $\left<M\right>$ of the distribution. Using the amplitude of fluctuations at the CMB, the amplitude of fluctuations at $a_{fct}$ reads,
\begin{equation}
    \left<|\delta|^2\right>^{1/2}_{a_{fct}}=\left<|\delta|^2\right>^{1/2}_{a_{cmb}}\frac{a_{eq}}{a_{cmb}}\left(\frac{a_{fct}}{a_{eq}}\right)^2\left(\frac{k(\left<M\right>)}{k_{piv}}\right)^{\frac{(n_b-n_s)}{2}},
    \label{eq: typical delta FCT}
\end{equation}
where $a_{eq}\simeq 2.94\times 10^{-4}$ and $a_{cmb}\simeq 9.08\times 10^{-4}$ are the scale factors at equality and CMB respectively, and we use $\left<|\delta|^2\right>^{1/2}_{a_{cmb}}\sim 10^-3$ . Also, we take into account the growth factors during matter and radiation domination, and the effect of the blue index $n_b$, where $k(\left<M\right>)$ is given by Eq. \eqref{eq: k_R(M) definition}.}

\JS{Instead, for the HC scenario the scale of interest is the horizon mass at the mean formation scale factor $\left<a\right>$ of the distribution,
\begin{equation}
    \left<|\delta|^2\right>^{1/2}_{\left<a\right>}=\left<|\delta|^2\right>^{1/2}_{a_{cmb}}\frac{a_{eq}}{a_{cmb}}\left(\frac{\left<a\right>}{a_{eq}}\right)^2\left(\frac{k(\left<a\right>)}{k_{piv}}\right)^{\frac{(n_b-n_s)}{2}}.
    \label{eq: typical delta HC}
\end{equation}
In this scenario, $k(\left<a\right>)$ is given by Eq. \eqref{eq: k_R(M) definition}, with $R$ as the Hubble radius evaluated at $\left<a\right>$.}

Notice that since the typical delta $\left<|\delta|^2\right>^{1/2}$ depends on $\left<M\right>$ and $\left<a\right>$ in each scenario respectively, its value will depend on the mass function parameters ($n_b,M_*$).

With $\left<|\delta|^2\right>^{1/2}$ in hand we can calculate the ratio $\delta_c/\left<|\delta|^2\right>^{1/2}$ which should be of order unity\JSH{, or at least less than unity}, as explained earlier. If $\delta_c \gg \left<|\delta|^2\right>^{1/2}$ almost none of the density fluctuations would collapse into a PBH \JSH{and Press-Schechter will not make sense anymore}. Conversely, \JSH{ when $\delta_c \ll \left<|\delta|^2\right>^{1/2}$ some relation with the primordial power spectrum is still preserved.}

\JS{\subsection{Details on the $f_m$ value}
\label{sec:app_fm}}




\JS{When choosing the minimum possible value of $f_m$ one  considers that all regions will form a PBH but only a fraction $f_m$ of the energy density of each region will actually collapse into it. The minimum possible value of $f_m$, is defined as the  value for which all linear overdensities with $\delta > \delta_c$ are associated with the formation of a PBH. This condition is accomplished when $f_m = \beta = \rho_{PBH}(a)/\rho(a)$ (see Eq \eqref{eq:beta_def}).}

\JS{Given that $\rho_{PBH}$ depends on the \NP{(mean) formation time, and therefore on} $f_m$ itself, obtaining this value is an iterative process, beginning with the assumption that $f_m = 1$ and then computing $\beta$, updating the $f_m$ value with the resulting $\beta$ value until convergence on $\beta$ is reached.}


\JS{As indicated in Section \ref{sec:app_ratiodeltas}, a plausible scenario for PBH formation requires $\delta_c/\left<|\delta|^2\right>^{1/2}\sim 1$. In our formalism, this translates into finding the right parameters that lead to this result. }

\JS{Then, we are interested in values of $f_m$ that favour  $\delta_c/\left<|\delta|^2\right>^{1/2}\sim1$. To achieve this, we first calculate the value of $\left<|\delta|^2\right>^{1/2}(f_m,n_b,M_*)$ (see Eqs \eqref{eq: typical delta FCT} and \eqref{eq: typical delta HC}). Then we take the expressions for $\delta_c(f_m,n_b,M_*)$ (Eqs \eqref{eq:Apdelta_c FCT brk} \eqref{eq: AP Mstar HC brk} for FCT and HC respectively) and solve
\begin{equation}
    \delta_c(f_m,n_b,M_*) = \left<|\delta|^2\right>^{1/2}(f_m,n_b,M_*)
    \label{eq: Appendix fm such that dc_dt is one}
\end{equation}
for $f_m$.}

\subsection{Fixed conformal time Mass Function}

In this scenario, the background energy density is given by

\begin{equation}
    \rho_{fct} = \left(\frac{\rho_{DM,0}}{a_{fct}^3}+\frac{\rho_{r,0}}{a_{fct}^4}\right),
    \label{eq: rho_fct}
\end{equation}
where $\rho_{DM,0}$, $\rho_{r,0}$ are the $z=0$ energy densities for dark matter and radiation respectively and $a_{fct}$ is the scale factor of PBH formation. Although in this epoch the matter contribution can be neglected because $\rho_{DM}(a_{fct}) \ll \rho_{r} (a_{fct})$, we have included it in our analysis. It should also be noted that the energy density for radiation does not include the neutrino contribution.
Then, we can directly derive the radius associated to a certain wavenumber $k_R$, using equation \eqref{eq: Horizon Mass}

\begin{equation}
    R(M) = \frac{1}{a_{fct}}\left(\frac{3}{4\pi \rho_{fct}}\right)^{1/3}\,\left(\frac{M}{f_m}\right)^{1/3},
\end{equation}
\noindent
with this, $k_R$ can be written as

\begin{equation}
    k_R(M) = C_{fct} \JS{\left(\frac{f_m}{M}\right)^{1/3}},
    \label{eq:k_R__fct}
\end{equation}
\noindent
where 
\begin{equation}
C_{fct} = a_{fct}\left(\frac{32\pi^4\rho_{fct}}{3}\right)^{1/3}.
\end{equation} 

\noindent
Note that $M$ corresponds to the mass of the PBH since we included the factor $f_m$ already. These are the necessary considerations for the FCT scenario. Now, things will be different when considering a standard PPS or a broken PPS due to the particular results on the calculation of $\sigma(M)$ (see eq. \eqref{eq:sigmasquared}). 

In the construction of the mass function, we define a characteristic mass scale $M_*$ that satisfies $\nu(M_*)\equiv 1$ and, since $\nu(M)$ depends directly on $\sigma(M)$, this parameter will be different for the two PPS considered \JS{ and will also depend on $f_m$}.
In the mass function this parameter $M_*$ is the mass where an exponential cut-off starts. This can be thought of as the characteristic mass in our mass distribution and, it is directly related with the \JS{linear} critical density contrast $\delta_c$.

\subsubsection{Standard Power Spectrum}

In this situation, the characteristic mass (see Appendix \ref{sec: appendix FCT} for further details of the derivation) is given by

\begin{equation}
    M_* = \JS{\left(\sqrt{\frac{4\pi\, a^4_{fct}\,(A_{s}/k_0^{n_s}) C_{fct}^{n_s + 3}}{n_s + 3}}\right)^{\frac{6}{(n_s +3)}}\,\frac{f_m}{\delta_c^{\frac{6}{(n_s +3)}}}}.
\end{equation}

\noindent
The mass function for a standard PPS is then given by

\begin{equation}
    \begin{split}
        \left(\frac{dn}{dM}\right)^{\text{std}}_{\text{fct}} = A_n\, \frac{\rho_{DM}(a)}{\sqrt{2\pi}}\, \frac{n_s + 3}{3\, M^2} \left(\frac{M}{M_*}\right)^{\frac{n_s +3}{6}} \, \\ \times\exp{\left[-\frac{1}{2}\left(\frac{M}{M_*}\right)^{\frac{n_s +3}{3}}\right]}.
    \end{split}
    \label{eq:mass_function_standard_fct}
\end{equation}

\subsubsection{Broken Power Spectrum}

The first difference that appears in this scenario is that we have an extra scale that corresponds to the pivot wavenumber $k_{piv}$. This implies that there is a particular mass $M_{piv} = (C_{fct}/k_{piv})^3 \JS{f_m}$ above which $P(k)$ corresponds to Eq. \eqref{eq:pkplaw}. In the regime of PBHs with masses below $M_{piv}$, one analogously obtains \footnote{the derivation of $M_{*}$ is given in the appendix \ref{sec: appendix FCT}}

\begin{equation}
    M_{*}\equiv \left(\frac{\delta_{c}^{2}}{\JS{f_m^{3\alpha}}A_{piv}\,S_{2}}-\frac{S_{1}}{S_{2}\JS{f_m^{\alpha}}}\right)^{\frac{-1}{\alpha}},
\end{equation}
\noindent
where $\alpha \equiv \frac{n_{b}+3}{3}$ and 

\begin{equation}
A_{piv} \equiv \frac{4\pi\,a^4_{fct}\,(A_{s}/k_0^{n_s})\, C_{fct}^{(n_{b}+3)}\, k_{piv}^{n_{s}-n_{b}}} {(n_s+3)(n_{b}+3)},
\label{eq:Apiv_fct}
\end{equation}

\begin{align}
    &S_1 \equiv (n_{b} - n_{s})\JS{\left(\frac{C_{fct}}{k_{piv}}\right)^{-3\alpha}},\nonumber\\
    &S_2 \equiv (n_s+3).
    \label{eq:S1 and S2 FCT}
\end{align}

\noindent
With this, the broken PPS mass function on the FCT scenario reads,

\begin{equation}
    \begin{split}
        \left(\frac{dn}{dM}\right)^{\text{brk}}_{\text{fct}} = A_n\frac{S_2\,\alpha\,\rho_{DM}(a)}{\sqrt{2\,\pi}\,M^{\alpha+2}} \frac{\left(S_1\JS{f_m^{-\alpha}} + S_2 M_{*}^{-\alpha}\right)^{1/2}}{\left(S_1\JS{f_m^{-\alpha}} + S_2 M^{-\alpha}\right)^{3/2}} \\
        \times\exp{\left[-\frac{S_1\JS{f_m^{-\alpha}} + S_2 M_{*}^{-\alpha}}{2\,\left(S_1\JS{f_m^{-\alpha}} + S_2 M^{-\alpha}\right)}\right]}.
    \end{split}
    \label{eq:mass_function_broken_fct}
\end{equation}


As expected, if we choose $n_{b} = n_s$ we recover the expression for the standard PPS eq. \eqref{eq:mass_function_standard_fct}. This also happens if we consider PBHs with $M>M_{piv}$ as a result of the existence of this pivot scale. Then, the final and most general expression for the FCT PBH mass function is

\begin{equation}
    \left(\frac{dn}{dM}\right)_{\text{fct}}= \begin{cases}
    \left(\frac{dn}{dM}\right)^{\text{brk}}_{\text{fct}} &\text{for $M<M_{\text{piv}}$},\\
    \\
    \left(\frac{dn}{dM}\right)^{\text{std}}_{\text{fct}} &\text{for $M\geq M_{\text{piv}}$},
    \end{cases}
    \label{eq:dndmFCT}
\end{equation}

\subsection{Horizon crossing Mass Function}

In this scenario, we are considering that \JS{the relevant amplitudes of linear fluctuations that can be linked to the physical formation of PBHs are restricted to the epoch of radiation domination.} Under this prescription, the energy density is

\begin{equation}
    \rho_{hc} \simeq\frac{\rho_{r,0}}{a_{hc}^4} \simeq \frac{3\,H_0^{2}}{8\,\pi\,G}\frac{\Omega_{r,0}}{a_{hc}^{4}}.
    \label{eq:energydensity_hc}
\end{equation}

Since we are now considering that the size of the \JS{linear} fluctuation matches the Horizon radius, we need to take into account that

\begin{equation}
    R(a_{hc}) = \frac{c}{a_{hc}H(a_{hc})} = \frac{c\,a_{hc}}{H_0\,\sqrt{\Omega_{r,0}}},
    \label{eq:Horizon_radius_hc}
\end{equation}
\noindent
where we have considered $H(a_{hc}) = H_0 \sqrt{\Omega_{r,0}} a_{hc}^{-2}$ of radiation domination. Thus the mass of the fluctuation using Eq. \eqref{eq: Horizon Mass} reads

\begin{equation}
    M(a_{hc}) =\frac{c^{3}a_{hc}^{2}}{2\,G H_{0}\sqrt{\Omega_{r,0}}} f_{m},
    \label{eq:M_hc}
\end{equation}
\noindent
where we used equations \eqref{eq:energydensity_hc} and \eqref{eq:Horizon_radius_hc}. This last expression gives a relation between the scale factor, $a_{hc}$, at which a mode of Lagrangian mass $M$ enters the horizon as

\begin{equation}
    a_{hc}=\left( \frac{2\,G H_{0}\sqrt{\Omega_{r,0}}}{c^{3}} \right)^{1/2}\left(\frac{M}{f_m}\right)^{1/2},
    \label{eq:a_hc}
\end{equation}
\noindent
allowing us to express $k_R$ as a function of the mass of the PBH

\begin{equation}
    k_{R}= C_{hc}\,\JS{\left(\frac{f_m}{M}\right)^{1/2}},
    \label{eq:kR_hc}
\end{equation}
\noindent
with $C_{hc}$ defined as
\begin{equation}
C_{hc} = \pi \left(\frac{2H_{0}\sqrt{\Omega_{r,0}}c}{G}\right)^{1/2}.     
\end{equation}

It is noteworthy that several authors give the relation of the PBH mass (or wavenumber) to the horizon mass in terms of the number of degrees of freedom of relativistic species at a certain epoch \citep[e.g.][]{Green:1997PhRvD,Nakama:2017,Inomata:2018,Gow:2020a}. Here, we only consider radiation in Eq. \eqref{eq:energydensity_hc} and the contributions of neutrinos, \JS{and other relativistic species} are neglected \JS{in the construction of the mass function}.

Just as we did above, we write our results for the mass function in terms of $M_*$. The meaning of this quantity remains the same but its relation with $\delta_c$ is different. 

\subsubsection{Standard Power Spectrum}

For this kind of power spectrum, the computation of $\sigma(M)$ (see the full derivation in Appendix \ref{sec:appendix HC}) leads to 

\begin{equation}
    M_* = \left[ \sqrt{\frac{4\,\pi\,(A_{s}/k_0^{n_s})}{n_s+3}} \left(\frac{G}{\pi\, c^2 }\right)^2 C_{hc}^{\frac{n_s+7}{2}}\right]^{\frac{4}{n_s-1}}\JS{\frac{f_m}{\delta_c^{\frac{4}{n_s-1}}}}.
    \label{eq: Mstar_hc_Standard_PPS}
\end{equation}
\noindent
This translates into the mass function

\begin{equation}
    \begin{split}
        \left(\frac{dn}{dM}\right)^{\text{std}}_{\text{hc}}=A_n\frac{ \rho_{DM}(a)}{\sqrt{2\pi}} \frac{(n_s-1)}{2} \frac{1}{M^{2}} \left(\frac{M_{*}}{M}\right)^{\frac{1-n_s}{4}} \\ \times\exp{\left[-\frac{1}{2}\left(\frac{M_*}{M}\right)^{\frac{1-n_s}{2}}\right].}
    \end{split}
    \label{eq:Mass_function_hc_Standard_PPS}
\end{equation}

\noindent
Note that for this mass function we need $n_s > 1$  in order to avoid a negative or null mass function. This issue has already been addressed by \citet{Carr:1994ar, Kim:1996PhRvD, Green:1999PhRvD, Chisolm:2006,Young:2014J, Gupta:2018}, among others. Considering that $n_s = 0.9649\pm 0.0042$ as measured by the Planck collaboration, we also explore the possibility of a broken PPS. 

\subsubsection{Broken Power Spectrum}

As before, in this scenario, the wavenumber $k_{piv}$ translates into a particular mass defined as $M_{piv} = (C_{hc}/k_{piv})^2\JS{\,f_m}$ due to the relation between $M$ and $k$ of Eq. \eqref{eq:kR_hc}. Then, $M_*$ is defined (further details of its derivation are given in appendix \ref{sec:appendix HC}) by

\begin{equation}
    \delta_c^2 = A'_{piv}\, \JS{f_m^{\frac{n_b-1}{2}}} \left[S'_1\,\JS{f_m^{-\alpha'}} M_*^2 + S'_2\, M_*^{2-\alpha'}\right],
    \label{eq:M_star_broken_hc}
\end{equation}
\noindent
where in this scenario we defined $\alpha' = \frac{n_{b}+3}{2}$,

\begin{equation}
    A'_{piv} \equiv  \frac{4\pi\,(A_{s}/k_0^{n_s})}{(n_s+3)(n_{b}+3)}\,\left(\frac{G}{\pi\, c^2}\right)^4\, \JS{\frac{C_{hc}^{(n_{b}+7)}}{k_{piv}^{n_b - n_s}}},
    \label{eq: Apiv_hc}
\end{equation}
and
\begin{align}
    &S'_1 \equiv (n_{b} - n_{s})\JS{\left(\frac{C_{hc}}{k_{piv}}\right)^{-2\alpha'}},\nonumber \\
    &S'_2 = S_2 \equiv (n_s+3).
    \label{eq: S1 and S2 HC}
\end{align}

\noindent
Note that, in this scenario, if we want to find $M_*$
 from a certain value of $\delta_c$ we need to solve a transcendental equation. Then Eq. \eqref{eq:M_star_broken_hc} is solved numerically for $M_*$. We can finally express our mass function for the broken PPS as 

\begin{figure}
    \centering
    \includegraphics[width=0.45\textwidth]{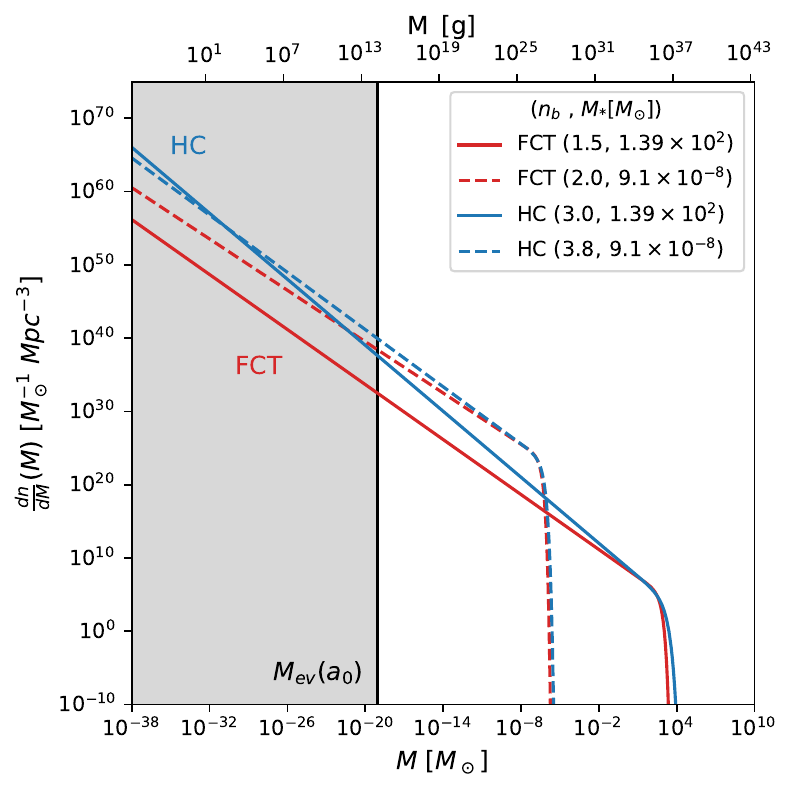}
    \caption{PBH mass functions for different scenarios, computed considering different values for $n_{b}$ and $M_{*}$. The red and blue lines correspond to the FCT and HC formation scenarios, respectively. The grey area indicates the PBHs which have already evaporated by $z=0$.}
    \label{fig:mass_functions}
\end{figure}

 \begin{equation}
 \begin{split}
     \left(\frac{dn}{dM}\right)^{\text{brk}}_{\text{hc}} = A_n\frac{ \rho_{DM}(a)}{\sqrt{2\pi}}\left[\left(\alpha'-2\right)S'_2\, M^{-\alpha'}- 2\,S'_1 \JS{f_m^{-\alpha'}}\right]\\
     \times \frac{\left[S'_1\,\JS{f_m^{-\alpha'\,}} M_*^2 + S'_2\, M_*^{2-\alpha'}\right]^{1/2}}{\left[S'_1\,\JS{f_m^{-\alpha'}}\, M^2 + S'_2\, M^{2-\alpha'}\right]^{3/2}}\\
     \times\,\exp{\left[-\frac{1}{2} \frac{\left[S'_1\,\JS{f_m^{-\alpha'}}\, M_*^2 + S'_2\, M_*^{2-\alpha'}\right]}{\left[S'_1\,\JS{f_m^{-\alpha'}}\, M^2 + S'_2\, M^{2-\alpha'}\right]}\right]}.
 \end{split}
 \label{eq:mass_function_broken_hc}
 \end{equation}
 
\noindent
Here we can set $n_s$ to the Planck value, however, the restriction will be on the blue spectral index $n_b$, requiring that $n_b > 1$. Once again, considering $n_b = n_{s}$ or $M > M_{piv}$, we recover the same expression as for the standard PPS (eq. \eqref{eq:Mass_function_hc_Standard_PPS}). 
It is also worth mentioning that $\sigma(M)$
must be a decreasing function in order  to be consistent with the cosmological
principle, i.e., larger scales are more homogeneous.
A decreasing $\sigma(M)$ has a negative derivative implying that the mass function is positive.
For higher masses $M>M_{piv}$, $\sigma(M)$ is an increasing function which does not satisfy the cosmological principle. To restore its consistency, we modify $\sigma(M)$ by considering that it has a constant minimum value for $M \geq M_{piv}$, implying that $\left(dn/dM\right)_{\text{hc}}=0$ for these masses.
Therefore, our final definition for the HC mass function is 

\begin{equation}
    \left(\frac{dn}{dM}\right)_{\text{hc}}= \begin{cases}
    \left(\frac{dn}{dM}\right)^{\text{brk}}_{\text{hc}} &\text{for $M<M_{\text{piv}}$},\\
    \\
    0 &\text{for $M\geq M_{\text{piv}}$}.
    \end{cases}
    \label{eq:dndmHC}
\end{equation}
\noindent
This functional form imposes an upper limit for the mass of the PBHs under this formation mechanism.  

It is relevant to mention that in Eq. \eqref{eq:energydensity_hc} we are using an approximation when we assume that the background density is only composed of radiation. This consideration only holds until matter-radiation equality $z_{eq}$. Therefore, the mass function obtained in \eqref{eq:dndmHC} is only valid \JS{for PBHs with $M\lesssim4,7\times10^{17}M_{\odot}$ (imposing the same limit on the $M_*$ value), linked to the amplitude of the linear fluctuation at $z_{eq}$}. For the rest of this work, we consider \JS{PBH formation related to linear fluctuations that enter the horizon up to $z_{eq}$}. 

\JS{
\subsection{PBH mass function examples}}

We show examples of these mass functions for different scenarios in Figure \ref{fig:mass_functions}. The red and blue lines shows the FCT and HC scenarios for two sets of parameters each one. In both, the solid line, corresponds to a higher $M_{*}$ value, predicting PBHs with high mass in contrast with the dashed line. Also, for both scenarios, each line has a different slope, related to the distinct $n_{b}$ values. The grey area, shows the region of PBHs evaporated today. We also show the values of $\left<M\right>$ for some parameter choices in the HC and FCT scenarios in Table \ref{tab:values_mean_mass}. This shows that the mean mass of the PBH distribution  is larger for larger $n_b$ values, approaching the value of $M_*$. This behaviour is expected as $n_b$ is related to the mass function slope.  Mass functions with low $n_b$ values have steeper slopes giving more weight to the low mass population hence decreasing the resulting mean mass. It is worth to note that this behaviour becomes more relevant in the HC scenario, where the resulting mass functions are steeper for the same $n_b$ value compared to the FCT scenario.

    \begin{table}
\caption{Mean mass of the PBH distribution for different parameters and the HC and FCT scenarios.}
    \centering
    \begin{tabular}{ccc}
    \hline
    $n_b$ & \multicolumn{2}{c}{$\left<M\right> \left[M_\odot\right]$}\\
    \hline
    & $M_* = 1.39\times 10^2 M_\odot$ & $M_* = 9.1\times 10^{-8} M_\odot$\\
    \cline{2-3}\\
    \multicolumn{3}{c}{\textbf{HC}}\\
    $1.1$ & $ 2.02\times 10^{-18} $ & $ 8.13\times 10^{-19} $\\
    $2.0$ & $ 2.22\times 10^{-14} $ & $ 1.28\times 10^{-16} $\\
    $3.0$ & $ 2.20\times 10^{-9 } $ & $ 6.23\times 10^{-14} $\\
    $4.0$ & $ 2.12\times 10^{-4 } $ & $ 2.89\times 10^{-11} $\\
    \multicolumn{3}{c}{\textbf{FCT}}\\
    $1.1$ & $ 1.03\times 10^{-5 } $ & $ 5.83\times 10^{-12} $\\
    $2.0$ & $ 8.40\times 10^{-3 } $ & $ 1.95\times 10^{-10} $\\
    $3.0$ & $ 3.64 $                & $ 4.14\times 10^{-9 } $\\
    $4.0$ & $ 2.61\times 10^{  1} $ & $ 1.71\times 10^{-8 } $\\
    \hline
    \end{tabular}
    \label{tab:values_mean_mass}
\end{table}

Since the broken PPS mass function can reproduce the standard PPS mass function scenario, we will focus on the broken PPS mass function as a general way to express our results.
Notice that the mass functions defined in Eqs. \eqref{eq:dndmFCT} and \eqref{eq:dndmHC} have several free parameters. Hereafter, we assume fiducial values for the cosmological parameters as mentioned above and $k_{piv}=10 \text{Mpc}^{-1}$ leaving as free parameters $n_b$ and $M_*$. We choose to fix $f_m=1$ in each scenario but also  \JS{adopt $f_m = \beta$  in some cases,} \JS{(See Section \ref{sec:app_fm})}. We consider only PBH mass functions that predict PBHs in the regime $M < M_{piv}$. This is reasonable, since $M_{piv} \gtrsim 10^{12}\, \text{M}_\odot$ for FCT and HC formation scenarios and PBHs  with higher masses tend to
be very rare.
We compute the slopes $n$ of the PS PBH mass distributions at masses
well below the characteristic mass $M_{*}$.
The slopes depend mainly on the blue spectral index. The logarithmic slopes of the differential mass functions are $n=-(9-n_{b})/6$ and $n=-(9-n_{b})/4$ for FCT and HC respectively.

\section{Constraints on the Fraction of Dark Matter in PBHs}\label{sec:constraints}

Once we already defined our different mass functions, we need to inspect \JS{the range of parameters of our modified PS scenario where} it is possible to account for all the dark matter in the form of PBHs\JS{, so as to provide a range of linear parameters that can be later tested against physical PBH formation mechanisms}.

\JS{The fraction of DM in the form of PBHs is usually expressed as} 

\begin{equation}
    f = \frac{\rho_{PBH}}{\rho_{DM}},
    \label{eq:fraction_of_DM_in_PBH}
\end{equation}
\noindent
being $f=1$ the scenario where PBHs can constitute all DM in the Universe. Bear in mind that PBHs form from primordial inhomogeneities which after the PBH formation continue evolving during radiation domination in a scale dependent way, thus giving rise to the transfer function.  This process reshuffles density fluctuations making PBHs essentially randomly distributed in space by the end of the epoch of radiation domination.  This implies that they can be simply considered ``very" cold dark matter (CDM) particles. Later on, during matter domination, when the density fluctuations grow into virialised structures, these are naturally formed by PBHs; i.e. dark matter haloes are PBH haloes.  Notice that just as in CDM, most of the PBHs live inside dark matter haloes (eg. \citealt[][]{Angulo2010})  

In the following, we investigate different constraints on this fraction.
In doing so we neglect the possibility that a fraction of the PBH population that collapsed to form dark matter haloes was expelled out of them due to two-body and other types of interactions. This is justified because massive PBHs decrease their potential energy to fall into the centre of the halo (and potentially merge) by kicking smaller PBHs out. To preserve the Virial equilibrium of the halo, the mass in PBHs kicked out must be of the order of the mass contributed by the most massive PBHs. This makes our approximation reasonable for steep mass functions where the mass in PBHs expelled outside the halo (only that of the massive ones) will be negligible compared to the total mass in PBHs in haloes. However, only in the case of the flattest mass functions (FCT scenario and large $n_b$ values), this  assumption may have some issues because in this case the  fraction of mass in PBHs expelled out of haloes may become significant.

\subsection{Constraint from super massive black holes mass function}\label{sec: SMBH}

The mass functions presented here allow for the existence of very massive PBHs along with a population of low mass PBHs, which does not occur with monochromatic PBH mass functions. Dark matter haloes collapse during matter domination from Lagrangian regions that contain PBHs already formed during the epoch of radiation.  Therefore, they contain PBHs drawn from the universal PBH mass function.  Depending on the volume that collapses, there is a maximum mass for the PBH that eventually falls into the dark matter halo, which we refer to as the central PBH in the Halo,

\begin{equation}
    V_\text{Halo}(M_{h})\,n_\text{PBH}(>M_c) = 1,
    \label{eq: Mcentral }
\end{equation}
where $V_\text{Halo}$ is the Lagrangian volume of the halo defined as

\begin{equation}
    V_\text{Halo}(M_h) = \frac{M_h}{\rho_{m}},
    \label{eq: Halo Volume}
\end{equation}

\noindent
with $\rho_{m}$ and $M_h$ corresponding to the comoving matter density and the halo mass respectively.  In this work we assume that the most massive PBH sinks to the center of the halo; we will also consider the possibility that it will merge with other PBHs later in this section.

\begin{figure}
    \centering
    \includegraphics[width=0.45\textwidth]{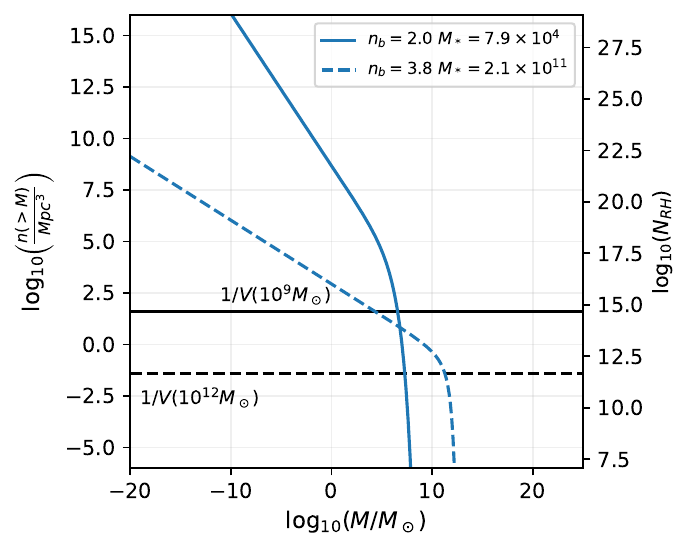}
    \caption{PBH cumulative number density \eqref{eq:cumulative number density} for two different sets of parameters $n_b$ and $M_*$, in the HC scenario, as a function of the PBH mass. The axis on the right also shows the logarithm of the number of PBHs within the comoving horizon. Black lines show the inverse of the Lagrangian volume of a halo with mass $10^{9}$ $M_\odot$ and $10^{12}$ $M_\odot$ as indicated.  }
    \label{fig: Cumulative mass function }
\end{figure}

If $M^*$ is large enough, a halo can contain central PBHs with masses exceeding even the largest known supermassive black holes (SMBH) observed in galaxies.  This would be at odds with observations and can be used to constrain the parameter space comprised by the $M_{*}$, $n_{b}$ parameters.  Figure \ref{fig: Cumulative mass function } shows examples of cumulative PBH mass functions with horizontal lines marking the inverse of the Lagrangian volume of dark matter haloes of different mass.  The point where these lines intersect the PBH mass functions show the central PBH mass that will be found in haloes of such mass, typically.  For example, the solid line indicates that both haloes represented in the figure  have $M_c \sim 10^{8}\,M_{\odot}$, while the dot-dashed line shows that, for those parameters, a halo with $M_h=10^{9}\,M_{\odot}$  has a central PBH with mass $M_c \sim 10^{5}\,M_{\odot}$and a halo with $M_h=10^{12}\,M_{\odot}$ has a central PBH of $M_c\sim 10^{11}\,M_\odot$.

\begin{figure}
    \centering
    \includegraphics[width=0.45\textwidth]{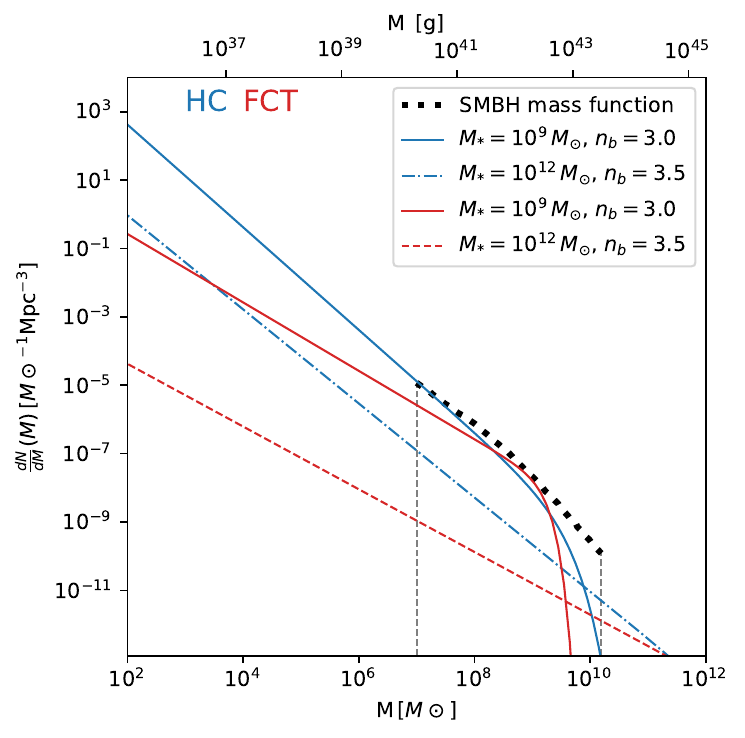}
    \caption{SMBHs mass function (dotted black line) with $\lambda=0.01$ \citep{Li:2012} in comparison with different PBH mass functions in the HC (blue) and FCT (red) scenarios.}
   \label{fig:dndmsmbh}
\end{figure}

The general idea of this constraint is that the abundance of massive PBHs should, in no case, surpass that of the SMBHs in galaxies. Then, the constraint will be built by comparing the abundance of the most massive PBHs in haloes with that of the SMBHs in galaxies, obtained from observations. The latter can be obtained from the active galactic nucleus (AGN) mass function as

\begin{equation}
\left(  \frac{dn}{dM}\right)_{SMBH} = \frac{1}{\lambda\left(  M\right)
}\left(  \frac{dn}{dM}\right)_{AGN} ,
\label{eq: SBMH mass function}
\end{equation}

\noindent
where $\lambda(M)$ corresponds to the duty cycle of AGNs, and the AGN mass function is that given by \citet{Li:2012}. For this work, we use $\lambda(M) = 0.01$, which is reasonable considering that we are interested in the SMBH mass function at $z=0$ \citep{Li:2012}. 
For the mass function of the most massive SMBHs in haloes we define the cumulative number density of SMBH as

\begin{equation}
n_{SMBH}\left(>  M\right)  =\int_{M}^{\infty}\left(  \frac{dn}{dM}\right)
_{SMBH}dM.
\end{equation}

Requiring that the PBH mass functions satisfy $n_\text{PBH}(M) < n_\text{SMBH}(M)$ (with $n_\text{PBH}$ defined as the cumulative number density of PBH) would be too restrictive as $n_\text{PBH}$ overestimates the number of PBHs because it also counts  satellite PBHs in the halo. We solve this by considering only the most massive PBH of mass $M_c$ within a halo, as defined above. 

We adopt the halo mass function proposed by \citet{Tinker:2008} to compute the cumulative number density of central PBHs

\begin{equation}
n_{\text{PBH}_{c}}\left(>  M\right)  =\int_{M}^{\infty}\left(  \frac{dn}{dM}\left(
M_{h}\left(  M_{c}\right)  \right)  \right)  _{\text{Halo}}\left(
\frac{dM_{h}}{dM_{c}}\right)  dM_{c},
\label{eq: cumulative central PBH}
\end{equation}

\noindent
where $M_h(M_c)$ is obtained implicitly through \eqref{eq: Halo Volume} and the Jacobian $\left(
\frac{dM_{h}}{dM_{c}}\right)$ is obtained using \eqref{eq: Mcentral }.

Figure \ref{fig:dndmsmbh} shows the observational SMBH mass function (dotted black line) in comparison with PBH mass distributions in the HC (blue lines) and FCT (red lines) scenarios assuming a $n_{b}=3.0$ and $3.5$ for $M_{*} =10^{9}  M_{\odot}$ and $ 10^{10} M_{\odot}$, respectively. Notice that for the latter $M_{*}$ values, the high mass tails for the PBH mass function are below the SMBH one at all SMBH masses, and are therefore allowed.

We  also study the effect of mergers of massive PBHs considering that PBHs with masses larger than the "sink-in mass" $M_{s}$ have all fallen to the centre and merged with the central PBH. $M_{s}$ is defined implicitly by requiring that the sink-in time into the centre of a halo of mass $M_{h}$ is equal to the Hubble time at redshift z. i.e.,

\begin{equation}
\tau_{\text{dyn}}\left(  M_{s},M_{h},z\right)  =H\left(  z\right)  ^{-1}.
\label{eq: dynamical time}
\end{equation}

The estimation of this dynamical time requires knowledge of the dynamical behavior of a massive object within a halo. This can be expressed in terms of the halo mass $M_{h}$ and the PBH mass $M_\text{PBH}$ \citep{Binney} as

\begin{equation}
\begin{split}
    \tau_{dyn}\left(  M_{PBH},M_{h},z\right)  = \frac{1.17}{\ln\left(
    \Lambda\left(  M_{PBH},M_{h}\right)  \right)  }\frac{r_{200}^{2}\left(
    M_{h},z\right)  v_{c}\left(  M_{h}\right)  }{GM_{PBH}},
\end{split}
\label{eq:tdyn full}
\end{equation}
where%
\begin{equation}
r_{200}\left(  M_{h},z\right)  =\left(  \frac{3M_{h}a^{3}\left(  z\right)
}{800\pi\rho_{m,0}}\right)  ^{\frac{1}{3}},
\end{equation}%
\begin{equation}
v_{c}\left(  M_{h}\right)  =\left(  \frac{GM_{h}}{r_{200}\left(  M_{h}\right)
}\right)  ^{\frac{1}{2}},%
\end{equation}
and%
\begin{equation}
\Lambda\left(  M_{PBH},M_{h}\right)  =1+\frac{M_{h}}{M_{PBH}}.
\end{equation}

\begin{figure}
    \centering
    \includegraphics[width=0.45\textwidth]{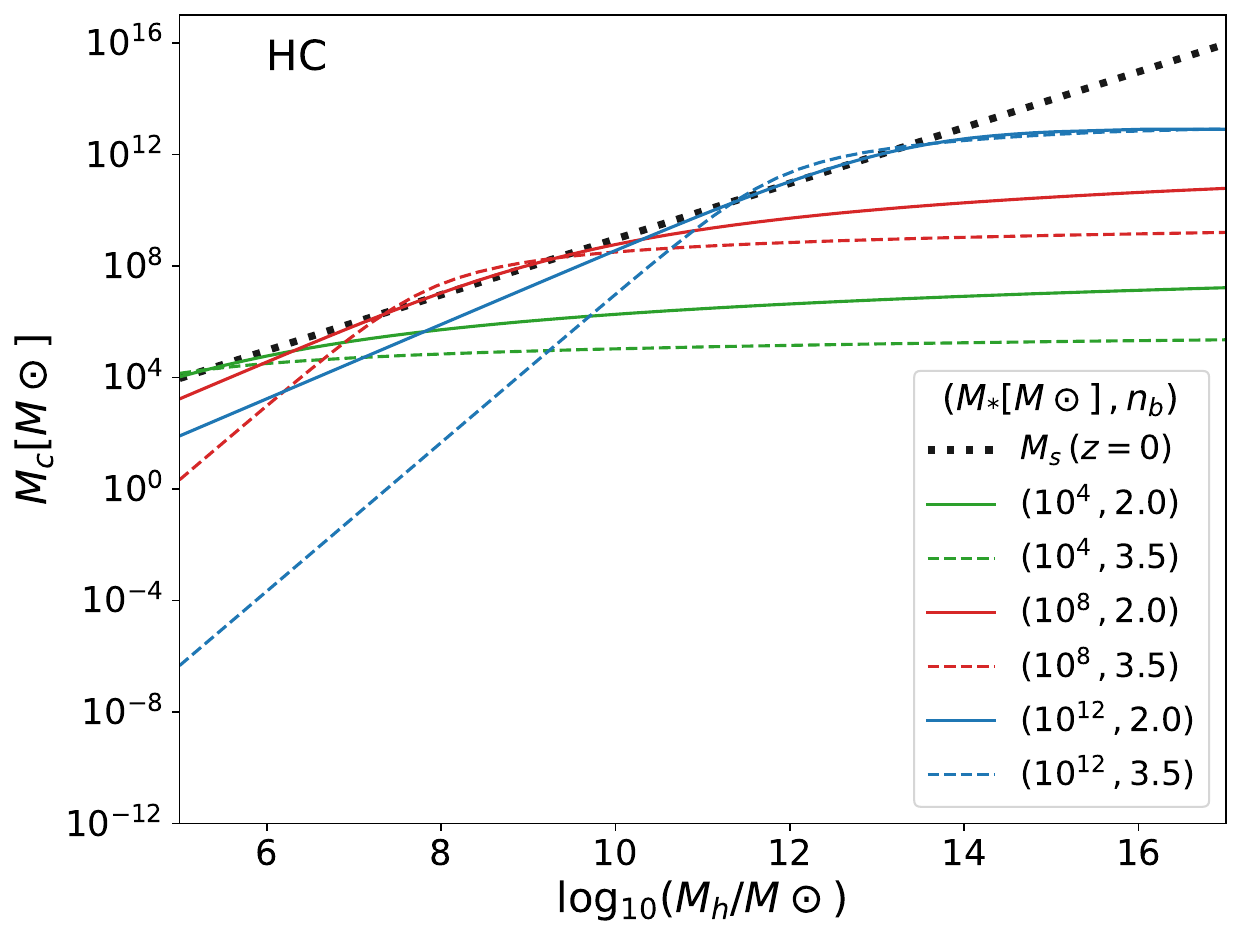}\\
    \includegraphics[width=0.45\textwidth]{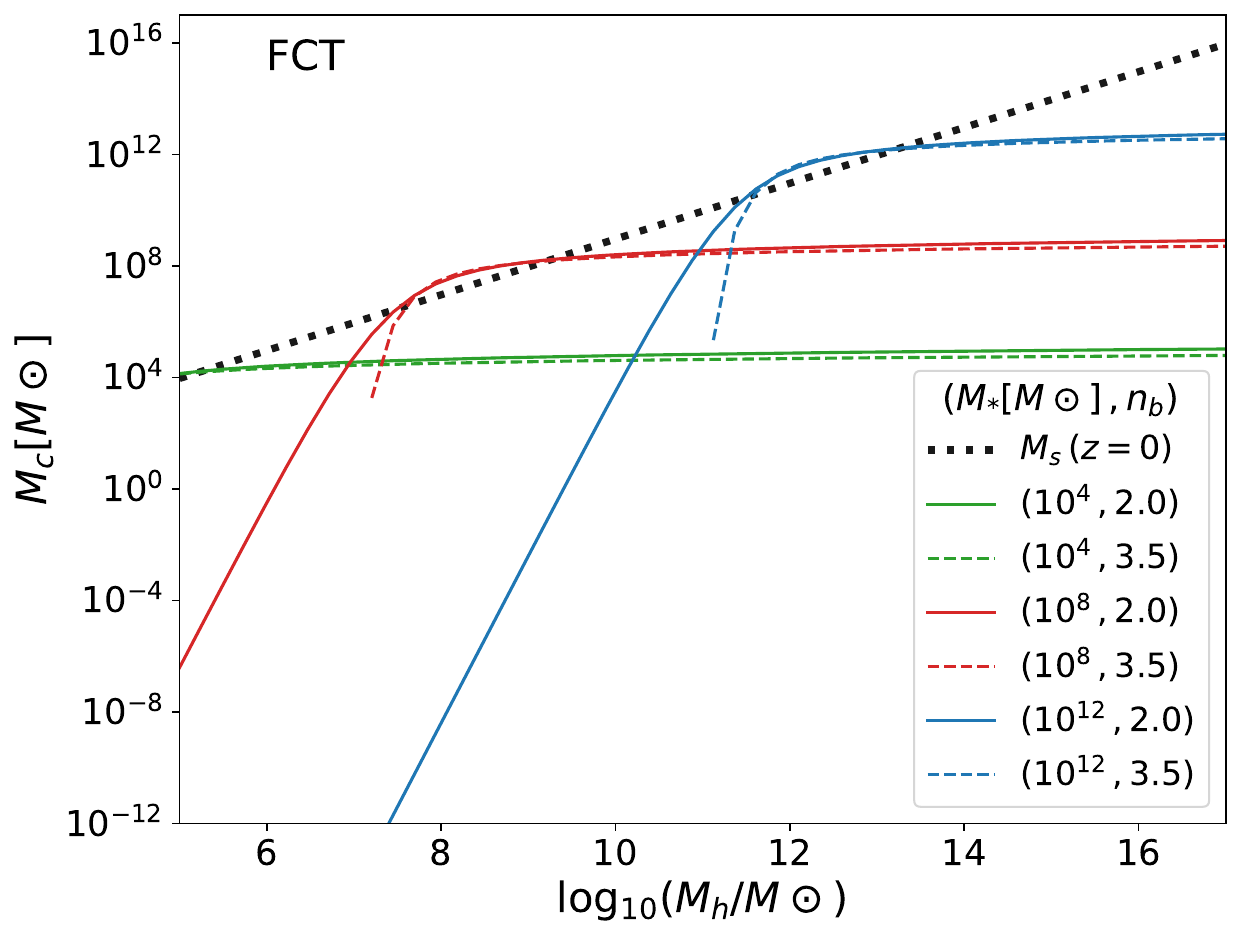}
    \caption{Sink-in mass ($M_{s}$, black dotted line) and central PBH mass ($M_{c}$) as function of the halo mass ($M_{h}$) for
    HC (top panel) and FCT (bottom panel) scenarios. The different colors correspond to different $M_{*}$ values with $n_{b}=2.0$ (solid lines) and $n_{b}=3.5$ (dashed lines). }
   \label{fig:smbh_C3}
\end{figure}

Then, we define the modified central PBH mass $M_{c}^{\prime}$ as
\begin{eqnarray}
&&M_{c}^{\prime}\left(  M_{h},z\right) = M_{c}\left(  M_{h}\right)\nonumber\\  
&&+V\left(
M_{h}\right)  \int_{\min\left(  M_{s}\left(  M_{h},z\right)  ,M_{c}\left(
M_{h}\right)  \right)  }^{M_{c}\left(  M_{h}\right)  }\left(  \frac{dn}%
{dM}\right)  _{PBH}MdM,\nonumber\\
\label{eq:mcprime}
\end{eqnarray}

\noindent
accounting for the merger of all PBHs with $M > M_{s}$ present in the halo, assuming instantaneous merging, and neglecting the satellite PBHs correctly. Also, in case that $M_{s}\geq M_{c}$, no PBH has the time to fall to the centre and the mass of the central PBH is not modified by merging. To investigate whether this merging effect is relevant we calculate the sink-in mass $M_{s}$ at $z=0$ and the central mass $M_{c}$ (Eq.\ref{eq: Mcentral }) in halos spanning a mass range of $10^{5}-10^{17} M_\odot$ for different PBH mass distributions. Figure \ref{fig:smbh_C3} shows the sink-in mass (dotted black line) and values of central mass as function of the halo mass. We show three $M_{*}$ values, $10^{4} M_{\odot}$ (green), $10^{8} M_{\odot}$ (red), $10^{12} M_{\odot}$ (blue) and two values for $n_{b}$ for each $M_{*}$, $n_{b}=2$ and $n_{b}=3.5$ represented by solid and dashed lines, respectively. As we can see, $M_{s}$ is greater than $M_{c}$ at almost all $M_{h}$ except in a small region when the halo mass is similar to the $M_{*}$ value. This means that, in general, the central PBHs in these halos are not modified by mergers $M_{c}^{\prime}\approx M_{c}$ \footnote{Notice that for FCT, there are no $M_{c}$ in halos with masses below $\sim 10^{7}M_{\odot}$ and $\sim 10^{11}M_{\odot}$ for $M_{*}=10^{8} M_{\odot}$ and $M_{*}=10^{12} M_{\odot}$ respectively when $n_{b}=3.5$.}. For the case when $M_{s}$ is similar to $M_{*}$, the mass of the formed halo should satisfy $M_{h} \gg M_{*}$.

\begin{figure}
    \centering
    \includegraphics[width=0.5\textwidth]{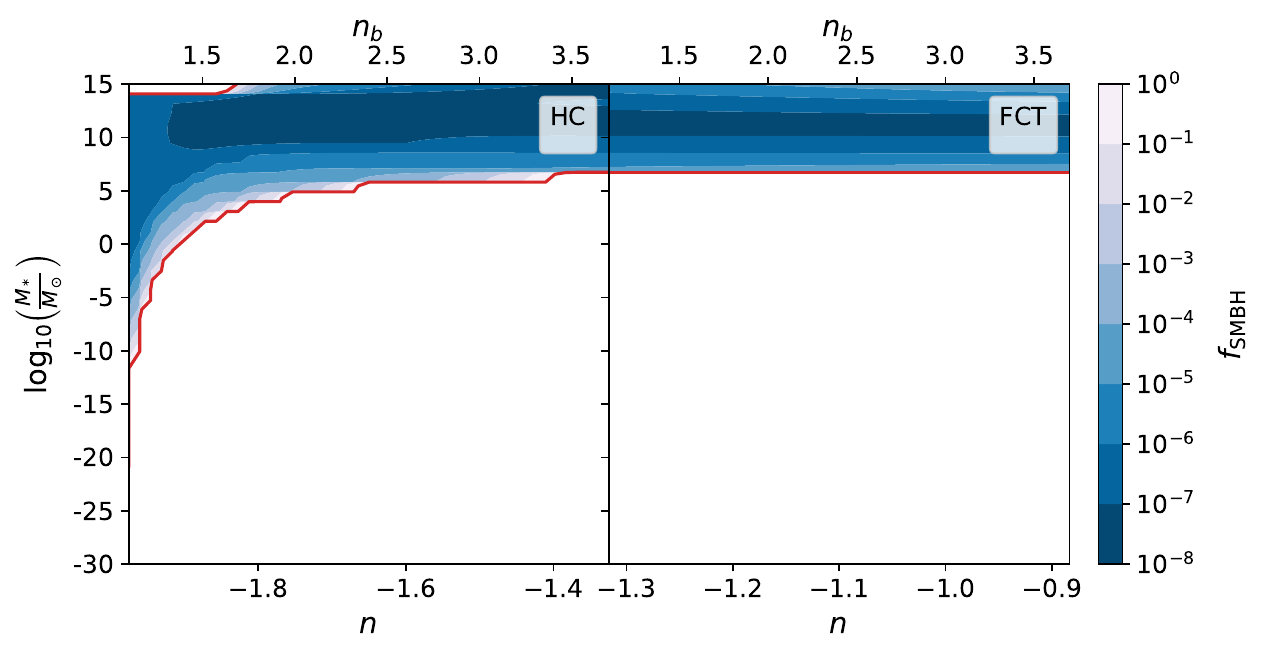}
    \caption{Level contours for $f$ given by Eq. \ref{eq:fSMBH} for HC (left panel) and FCT (right panel) scenarios. The red lines indicate the $f_{\text{SMBH}}=1$ contours. As the figure suggests, the regions where this constraint becomes more restrictive corresponds to the ones with high $M_{*}$, i.e., regions where their corresponding mass function predicts massive PBHs.}
    \label{fig:smbh_C2}
\end{figure}

Finally, it will suffice to impose that $n_{\text{PBH}_{c}}(>M) \leq n_{\text{SMBH}}(>M)$ for all values of $M$, considering the cumulative number density of central PBHs as in \eqref{eq: cumulative central PBH}. This translates into a permitted fraction of DM in the form of PBHs given by

\begin{equation}
    f  =\min\left( \left\{ \frac{n_{SMBH}\left( > M\right) } {N_{PBH_{c}} \left( >M\right)  }\right\}_{M>10^{7}M_{\odot}}\right),
    \label{eq:fSMBH}
\end{equation}

\noindent
where we consider the minimum value of this fraction in order to be in agreement with the SMBH mass function, even in the most restrictive scenario. Figure \ref{fig:smbh_C2} shows the contours of different values of $f$ in the $n$, $M_{*}$ and $n_{b}$ space for this criterion (Eq. \ref{eq:fSMBH}) for the FCT and HC scenarios. As we can see, this constraint affects the scenarios where the mass functions predict PBHs with high masses ($\mathcal{O}(10^{10}\, M_\odot)$) since these are the ones that can show disagreement with the observationally detected SMBHs. 

In the remainder of this section, we present the other constraints on the fraction of DM in PBH, extracted from the literature.

\subsection{Monochromatic Constraints and extended PBH mass distributions}\label{sec: MMD to EMD}

Most of the constraints for $f$ are computed for a monochromatic mass function. One would want to calculate again these constraints but now, considering that the primordial black holes span a wide range of masses. 
Nevertheless, the physical processes in most of these observable constraints are not completely understood 
and many astrophysical parameters have to be assumed. Thus, for extended mass distributions the computation of $f$ considering the mass dependence on the different physical processes becomes very difficult \citep{Carr:2017}. 

Some authors have developed methods to translate the constraints on monochromatic mass functions into extended ones \citep[see for instance][]{Carr:2017,Bellomo_2018}. Based on these approaches, we propose an alternative formalism to constrain PBH extended mass functions from monochromatic bounds. This new method accounts for the fact that the physical processes used to constrain $f$ are sensitive to PBHs only in a particular mass range, which accounts for a fraction of the total PBH mass in an extended mass function. It also accounts for the redshift evolution of the mass function due to PBH evaporation and the entrance of more massive PBHs to the Hubble volume as time progresses.

We consider different physical processes which independently provide
constraints to the allowed DM fraction in PBHs assuming monochromatic
distributions. Each underlying process is related to some observable output,
which is assumed to be extensive in the number of PBHs, i.e., the total output
is proportional to the number of PBHs. Then, the fraction $f$ is interpreted
directly as the ratio between the maximum allowed output and the measured
value of that output. If this ratio is greater than one it means that the
constraint, given by the maximum allowed value of the output, has not been
reached. If it is less than one, it measures the maximum fraction of DM that PBHs can account for, such that when multiplied by the total measured
value of the output, one recovers the maximum allowed value.

The inverse of $f$ thus gives the observed output normalised by its maximum
allowed value (which corresponds to the observational constraint), and is thus
a measure of the normalised output function $g\left(  M\right)  $. Due to the
extensive nature of the output, and considering that an extended mass function
for PBHs can be interpreted as the sum of different monochromatic populations,
the average of the output function $\left\langle g\left(  M\right)
\right\rangle $ can be calculated using the mass function itself as the
relevant statistical weight. This value of an effective output is interpreted
as the resulting output from a combination of BH populations of different
masses, with a distribution provided by the mass function,

\begin{equation}
\left\langle g(M)\right\rangle =  \frac{\int_{max(M_{min},M_{ev}(z))}^{min(M_{max},M_{1pH}(z))}g(M)\,\frac{dn}{dM}\,dM}{\int_{max(M_{min},M_{ev}(z))}^{min(M_{max},M_{1pH}(z))}\,\frac{dn}{dM}\,dM},
    \label{eq:Mean g}
\end{equation}
\noindent
where $M_{min}$ and $M_{max}$ are the mass limits where the observational constraint is sensitive, and we only consider PBHs such that they exist (e.g., they have not evaporated at that moment) within the causal volume by including $M_{ev}(z)$ and $M_{1pH}(z)$. Then, the multiplicative inverse of this $\left\langle g\left(  M\right)
\right\rangle $ is interpreted directly as the effective $f$ for the PBH
population, characterised by a particular choice of the mass function. This procedure is general, and it only relies on the assumption of extensivity of the underlying physical quantity associated with the observational constraint.

\begin{figure}
    \centering
    \includegraphics[width=0.42\textwidth]{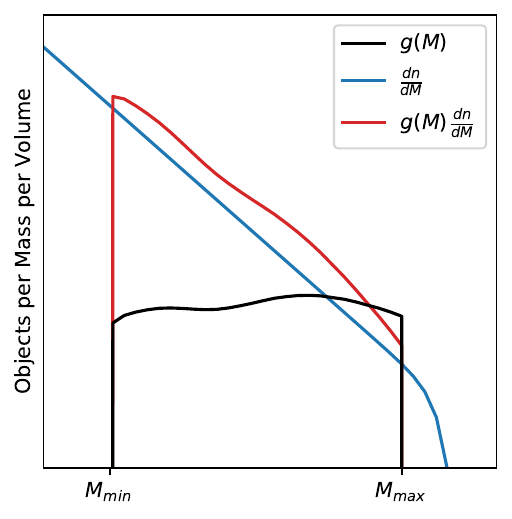}
    \caption{Schematic figure for the $C_M$ correction. The blue line corresponds to an arbitrary extended mass distribution. The black line represents the $g(M)$ function, which is only sensitive to masses within $(M_{\min},M_{\max})$ and the red line shows the mass function weighted by $g(M)$. This figure shows that when considering a particular $g(M)$, related to some constraint, it is possible to have masses outside the regime where $g(M)$ is sensitive.}
    \label{fig:Constraint_pop}
\end{figure}

Because the normalised output functions $g\left(  M\right)  $ have support only on a domain that is a subset of the considered mass range for PBHs where the mass function is defined, the effective fraction $f$ obtained as previously mentioned has to be corrected to account for the mass not constrained by $g\left(M\right)$. Figure \ref{fig:Constraint_pop} illustrates a generic mass function and a particular output function $g\left(M\right)$. For instance, consider two populations (A) and (B) of PBHs. The  population (A) corresponds to the PBHs that can affect the observable measured by a particular constraint. For example, if some constraint is sensitive to objects with masses between $M_{min}$ and $M_{max}$ then, only PBHs within those masses will be considered on the calculation of the effective $f$. The population (B) corresponds to the whole population of PBHs, considering even the ones that cannot be detected by this constraint. 
For instance, if the population (B) holds more PBH mass than population (A), then the resulting effective $f$ will be higher since the constraint will only act on a small fraction of our population and hence, a small fraction of the total PBH mass.

To correct for this, we define the constrained mass density as 

\begin{equation}
\rho_{constr} =  \int_{max(M_{min},M_{ev}(z))}^{min(M_{max},M_{1pH}(z))} \left(\frac{g(M)}{g_\text{max}}\right)M \frac{dn}{dM}dM,
    \label{eq: constrained mass density}
\end{equation}

\noindent
where $\frac{g(M)}{g_\text{max}}$ will act as a filter function, varying from 1, at the maximum value of $g(M)$, $g_\text{max}$, and 0, whenever M is outside the domain of $g(M)$. 

We use this definition to compute the ratio between the total mass density in PBHs at redshift $z$ (see eq. \eqref{eq:rho_pbh}) and the constrained mass density \eqref{eq: constrained mass density} at the same epoch, obtaining

\begin{equation}
    C_{M} (z)= \frac{\int_{M_{ev}(z)}^{M_{1pH}(z)} M \frac{dn}{dM}dM}{\rho_{constr}}.
    \label{eq:Cm}
\end{equation}

\noindent
The meaning of $C_{M}(z)$ is understood such that all the mass is constrained near the maximum of $g\left(  M\right) $, with the sensitivity of the constraint decreasing proportionally to the decrease in observable output away from the maximum $g_{max}$. 

\begin{figure}
    \centering
    \includegraphics[width=0.42\textwidth]{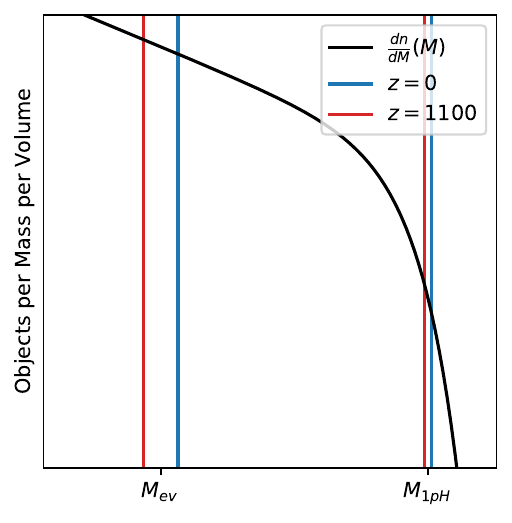}
    \caption{Schematic figure for the $C_z$ correction. The black solid line represents an arbitrary mass function. The red and blue vertical lines mark the evaporation ($M_{ev}$) and $M_{1ph}$ masses at $z=0$ and $z=1100$ respectively. Notice the number of PBHs in these boundaries change at different redshifts.}
    \label{fig:Cz}
\end{figure}

Additionally, the data to compute some of the observational constraints is obtained at a particular redshift $z$. Then, we need to introduce another correction to take into account the evolution of the mass function from this redshift $z$ to the current epoch. Figure \ref{fig:Cz} illustrates a mass function whose limits are the evaporation mass $M_{ev}$ and $M_{1pH}$ today and at $z=1100$. Notice that, within these limits, there is a difference in the mass function and mass density of PBHs when different redshifts are considered. Therefore, we introduce the $C_{z}$ quantity to correct the resulting effective $f$ by the mass function evolution as

\begin{equation}
C_{z}(z) =
      \frac{ \int_{M_{ev}(z=0)}^{M_{1pH}(z=0)} M \frac{dn}{dM}dM}{\int_{M_{ev}(z)}^{M_{1pH}(z)} M \frac{dn}{dM}dM}.
\label{eq:Cz}
\end{equation}

Thus, the corrected fraction of DM as PBHs is then given by the fraction computed as $\left\langle g\left(  M\right)  \right\rangle
^{-1}$, multiplied by the correction factors $C_{M}$ and $C_{z}$, i.e.,
\begin{equation}
    f_i= \left(\frac{C_{M}(z_i)\, C_{z}(z_i)}{\left\langle g_i\left(  M\right)  \right\rangle}\right),
    \label{eq: corrected fraction}
\end{equation}

\noindent
where the subscript $i$ was added to indicate that this is for a particular monochromatic constraint.

In the following, we apply this procedure to compute the allowed $f$ for HC and FCT extended mass distributions for a range of values for the parameters $M_{*}$ and $n_{b}$. We have chosen monochromatic constraints from evaporating PBHs, lensing and dynamical effects covering a wide range of masses; when there are multiple observables producing constraints on the same monochromatic PBH mass, we choose the more restrictive ones that span the widest range of masses, which are the most restrictive for extended mass distributions.  This ensures that our combined constraints will be complete and contain all relevant observables.   
For a set of parameters $(M_{*},n_{b})$, the maximum allowed fraction of DM in the form of PBHs $f$, will be the minimum $f_i$ from all considered constraints i.e.

\begin{equation}
f(M_{*},n_{b})=\text{min}(f_{i}).
\label{eq: minimum f}
\end{equation}

As a summary, the method to translate constraints on monochromatic mass distributions to constraints on extended mass distributions, described in this section, considers the following steps, for each monochromatic mass function constraint:

\begin{enumerate}
    \item If the analytic function of $f$ is not given, we obtain data points from the $f$ plot given in the literature.
    \item We compute $\left\langle g(M)\right\rangle$ using Eq. \eqref{eq:Mean g}.
    \item We calculate the two corrections $C_M$ and $C_z$ given by Eq. \eqref{eq:Cm} and Eq. \eqref{eq:Cz} respectively, taking into account the redshift of the observations $z_i$.
    \item The corrected $f$ is then calculated using Eq. \eqref{eq: corrected fraction}.
    \item Once we performed the previous steps for all the monochromatic mass distribution constraints, the resulting admitted fraction of DM in PBHs is given by the minimum fraction obtained Eq. \eqref{eq: minimum f}.
\end{enumerate}

It is worth to note that the observational constraints depend on different astrophysical assumptions and most of them have caveats on the black hole physics. Therefore, it is important to understand the physics behind each process
which we briefly describe below.

\subsubsection{Big bang nucleosynthesis (BBN)}
The effect of low mass PBH evaporation on the BBN epoch has been already studied in several works \citep{Miyama:1978,Vainer:1977,Zeldovich:1977, Lindley:1980, Keith:2020,Carr:2020arXiv200212778C}. 
The particles radiated by PBHs could affect the abundance of primordial light elements, for instance, enhancing the neutron-proton ratio, hence increasing the helium abundance. Also this radiation can break the Helium nuclei and decrease the amount of Deuterium at the moment of BBN. Here, we consider the measurements of the primordial mass fraction $Y$, the ratio $D/H$, $^{6}Li/^{7}Li$, $^{3}He/D$ which impose bounds on the $\beta$ parameter (Eq. \eqref{eq:beta_def}) presented by \citet{Carr:2010}. 

This parameter is associated to the current density parameter of non evaporated PBHs and therefore, related to $f(M)$ as

\begin{equation}
    f(M) \approx 3.81\, \times\, 10^{8} \beta'(M) \left(\frac{M}{M_{\odot}}\right)^{-1/2},
\end{equation}
where \JSH{this expression is given by equation (55) in \citep{Carr:2020arXiv200212778C} and $\beta'(M)$ is related to $\beta(M)$ through
\begin{equation}
\beta^{\prime}(M) \equiv \gamma^{1 / 2}\left(\frac{g_{* \mathrm{i}}}{106.75}\right)^{-1 / 4} \beta(M),
\label{eq: beta prime}
\end{equation}
where $\gamma$ is related to the physics of the gravitational collapse and $g_{*i}$ corresponds to the number of relativistic degrees of freedom which, contrary to $\gamma$, can be specified very precisely \citep[as explained in][]{Carr:2010}.}

\JSH{Finally, we use this $f(M)$ (for a monochromatic distribution) to obtain the effective $f$ with our method, considering} a redshift $z=10^{10}$ for these constraints.

\subsubsection{Extragalactic $\gamma$-ray background}
\label{subsec:extragammaray}
A primordial black hole with mass $M$ can emit thermal radiation through the Hawking radiation mechanism. The emission rate for particles with spin $s$ in the range of energies $[E, E+dE]$ has been calculated by many authors \citep{MacGibbon:1990, MacGibbon:1991,Carr:2016_gammaray}. This phenomenon can be used to constrain the fraction of PBHs by confronting the theoretical spectrum of radiation (photons) emitted from PBHs with observations, for instance, the diffuse extragalactic $\gamma$-ray background (EGB). Different experiments have measured the diffuse EGB in the energy range $1$MeV-$1000$MeV \citep[see ][and references therein]{Carr:2020arXiv200212778C}. The observed extragalactic intensity is $I^{obs}\propto E^{-(1+\epsilon)}$, where $\epsilon$ parameterises the spectral tilt. From this relation, it is possible to estimate the fraction of PBHs as DM as
\begin{equation}
    f(M)_{\gamma} \lesssim 2 \times 10^{-8} \left(\frac{M}{M_{\gamma}}\right)^{3+\epsilon}
\end{equation}
where $M>M_{\gamma}$, $M_{\gamma}\sim5\times 10^{14}$g ($\sim 2.51\times10^{-19} M_{\odot}$) and $\epsilon$ takes a value between $(0.1,0.4)$. For this constraint, we assume a redshift $z=1160$, a minimum mass $M_{min}=2.52\times 10^{-19}M_\odot$, $M_{max}=6.4\times 10^{-17}M_\odot$ and $\epsilon=0.2$. Additionally, we consider the constraint on $f$ when $M<M_{\gamma}$ in the interval $\sim 1.3\times10^{-20}M_{\odot}< M < 2.51\times10^{-19} M_{\odot}$ by taking data points from the $f$ plot by \citet{Carr:2020arXiv200212778C}.

\subsubsection{Galactic center $\gamma$-ray constraint}
The current observations of the $511$ keV gamma-ray line from the Galactic centre by the INTEGRAL observatory \citep{Siegert:2016} can be used to constrain the fraction of PBHs as dark matter with masses $M\sim 10^{-17} M_{\odot}$, as they could radiate positrons which eventually annihilate producing a $\gamma$-ray spectrum.
We extract the data points from the constraint on $f$ obtained by \citet[][see also, \citealt{Laha:2019PhRvL,DeRocco:2019PhRvL}]{Dasgupta:2019} using the INTEGRAL measurements in the interval $\sim 5.5 \times 10^{-19} M_{\odot}$ and $\sim 8.5 \times 10^{-17} M_{\odot}$. This constraint is relevant at $z=0$.

\subsubsection{Gravitational lensing constraints}
\label{subsec:lensing_constraints}

Microlensing is the effect of an amplification of a background source during a short period of time produced by the passage of a compact object close to its line of sight. \citet{Paczynski1986} suggested that a population of objects producing this effect could be detected within the Milky Way halo. Each microlensing event will occur when a compact object goes through what is called the microlensing ``tube'' which is directly related to the mass of the object, in our case, the PBH. The observable for this constraint is the number of observed events and, this is in turn related to the number density distribution as a function of mass, i.e., the mass function of the objects \citep{kim1991,DeRujula:1991}. The microlensing of stars in the Magellanic clouds by massive compact halo objects (MACHOs) has been used to test the fraction of DM as PBHs \citep{Paczynski1986} in the range $\sim 10^{-8} M\odot < M < 60 M\odot$. 
Other campaigns to search lensing events of sources in the  Magellanic clouds due to MACHOs are the EROS \citep{Hamadache:2006,Tisserand:2007} and OGLE \citep{Ogle:2011} experiments.
For these constraints we compute $\left\langle g\left(  M\right)  \right\rangle$ from the $f(M)$ functional form of the curve plotted in \citet{Carr:2016,Carr:2020arXiv200212778C}.
The range of masses that can be constrained are $5.8\times 10^{-8} M\odot \lesssim M \lesssim 5M\odot$ and $1.8\times 10^{-7} M\odot \lesssim M \lesssim 0.3M\odot$ for EROS and OGLE measurements, respectively.

We also consider the limits on the abundance of compact objects which could produce a millilensing effect of radio sources \citep{Wilkinson:2001PhRvL}. This observable puts constraints on $f$ in the interval $\sim 1\times 10^{5} M_{\odot}<M<1\times 10^{9} M_{\odot}$. The femtolensing effect of gamma ray burst (GRBs) by compact objects also imposes a limit on $f$ in the interval $5\times 10^{-17}M_{\odot}  \lesssim M \lesssim 1\times 10^{-14} M_{\odot}$ \citep{Marani:1999, Nemiroff:2001, Barnacka:2012}. Nevertheless \citet[][see also \citealt{Carr:2020arXiv200212778C}]{Katz:2018} claim that most of the GRBs are are inappropriate for femtolensing searches and hence $f$ is not robustly constrained. The microlensing search of stars in the Milky way and M31 by PBHs with the Subaru Hyper Suprime-Cam provides a bound on $f$ in the interval $3.6 \times 10^{-12} M_{\odot} \lesssim M \lesssim 6.8 \times 10^{-6} M_{\odot}$ \citep{Niikura:2019}. Recently, \citet{Smyth:2020} point out that these constraints assume a fixed source size of one solar radius. By performing a more realistic analysis, they conclude that the current bounds are weaker by up to almost three orders of magnitude.
All these constraints are at redshfit $z=0$.

\subsubsection{Neutron star capture and white dwarfs}
Another constraint on the fraction of DM as PBHs is obtained from their capture by neutron stars in environments with high density, such as, the core of globular clusters. If a neutron star captures a PBH it can be disrupted by accretion of its material onto the PBH. Thus, the observed abundance of neutron stars imposes constraints on $f$ at a certain range of masses. The $f(M)$ function encoding the physics of the capture probability by neutron stars is given by    

\begin{equation}
    f_{NS}(M)=\frac{M}{4.7\times 10^{24}\mathrm{g}}\left(1-\exp{\left(- \frac{M}{2.9\times 10^{23}\mathrm{g}}\right)} \right)^{-1},
\end{equation}
where we have adopted the same values as \citet{Capela:2013}. Notice that this constraint is valid in the mass range $1.25\times 10^{-15} M\odot<M< 5\times 10^{-9} M\odot$. Additionally, the possibility that PBHs can trigger white dwarf explosions as a supernovae also provides a bound on $f$ \citep{Graham:2015}. The white dwarf distribution imposes constraints to PBHs with masses between $\sim 1.2\times 10^{-15} M_{\odot}$ and $\sim 1.6 \times 10^{-11} M_{\odot}$.
Both constraints are relevant at $z=0$.

Note, however, that \citet{Montero-Camacho:2019} has recently stated that the NS constraint is no longer valid. One of the reasons is that it considers globular clusters as an environment with high DM density. This can happen if the clusters are from primordial origin, however this scenario is not fully determined. \citet{Montero-Camacho:2019} also studied NS capture considering the environment on dwarf galaxies concluding that the survival of stars cannot rule out PBHs as DM. They also conclude that there is not an effective constraint from white dwarf survival.

\subsubsection{X-ray binaries.}

We know that there is a possibility that a PBH can accrete baryonic matter from the interstellar medium (ISM), forming an accretion disk which can radiate. \citet[][see also \citealt{Carr:2020arXiv200212778C}]{Inoue:2017} considered that a PBH can accrete material through the Bondi-Hoyle-Lyttleton accretion. In this approach, the mass accreted onto a PBH can be converted in radiation, associated to a luminosity $L$, and hence, the number of accreting PBHs, emitting a luminosity $L_{x}$ (in X-rays), can be estimated. Indeed, the luminosity function of X-ray binaries (XRB) restricts the maximum X-ray output and hence, the number of accreting PBHs. This luminosity function has been obtained using Chandra observations \citep[see for instance][]{Mineo:2012}, with $L_{x}$ spanning the range $10^{35}-10^{41} \mathrm{erg\,s^{-1}}$, implying constraints on PBH with masses $\sim 5.7 M_{\odot}-2\times 10^{7} M_{\odot}$.

We translate this into our method by extracting data points from the $f_{XRB}(M)$ curve presented by  \citet{Inoue:2017, Carr:2020arXiv200212778C}. This constraint applies to redshift $z=0$.

\subsubsection{Disruption of globular clusters and galaxies}
Another important constraint comes from PBH dynamical effects on astrophysical systems like globular clusters (GC) and galaxies (G) \citep{Carr:1999ApJ_dynamical}. A passing PBH could disrupt a GC due to tidal forces, thus the GC survival imposes the following bound on $f$

\begin{equation}
f_{GC}(M)=  \begin{cases}
      (\frac{M}{3\times10^{4} M\odot})^{-1}\qquad 3\times 10^{4} M\odot < M < 10^{6} M\odot,\\
      0.03 \qquad 10^{6} M\odot < M < 10^{11} M\odot,\
    \end{cases} 
\label{eq:fgc}
\end{equation}
which is relevant at $z=0$. Besides individual PBHs, hypothetical clumps could also disrupt galaxies in clusters, resulting in an additional bound on $f$ given by

\begin{equation}
f_{G}(M)=  \begin{cases}
      (\frac{M}{7\times10^{9} M\odot})^{-1}\qquad 7\times 10^{9} M\odot < M < 10^{11} M\odot,\\
      0.05 \qquad 10^{11} M\odot < M < 10^{13} M\odot,\
    \end{cases}
\label{eq:fgal}
\end{equation}
where this is relevant at $z=1$.
\subsubsection{Disk heating}
PBH encounters with (mainly old) disk stars could be responsible for disk heating in galaxies \citep{Carr:1999ApJ_dynamical,Carr:2020arXiv200212778C}. This dynamical effect is translated into a restriction on $f$ for high mass PBHs as
\begin{equation}
f_{DH}(M)=  \begin{cases}
      (\frac{M}{3\times10^{6} M\odot})^{-1}\qquad 3\times 10^{6} M\odot < M < 3\times10^{9} M\odot,\\
      \frac{M}{M_{halo}} \qquad 3\times 10^{9} M\odot < M < M_{halo}, \
    \end{cases}  
\end{equation}
where a halo mass, $M_{halo}$, of $3\times 10^{12}M\odot$ is assumed
and it is considered to be important at redshift $z=1$.

\subsubsection{Wide binaries}
Binary star systems with wide separations could be disrupted by encountering PBHs \citep{Chaname:2004ApJ,Quinn:2009MNRAS}. Observations of wide binaries in the Milky way impose a constraint on $f$ as a function of the PBH mass, given by

\begin{equation}
f_{WB}(M)=  \begin{cases}
      (\frac{M}{500 M\odot})^{-1}\qquad 500 M\odot < M \lesssim 10^{3} M\odot,\\
      0.4 \qquad 10^{3} M\odot \lesssim M < 10^{8} M\odot.\
    \end{cases}
\label{eq:fWB}
\end{equation}
This constraint is relevant at $z=0$.

\begin{table*}
 \caption{Summary of all the constraints considered in this work. We include the respective mass regimes where they apply and the corresponding redshift for each one, along with the mass of the PBH that ends its evaporation at that redshift $M_{ev}(z)$. Also, the last column shows  references which explain these constraints in more detail. Constraints marked with $^*$ are considered disputed for different reasons.}
    \centering
    \begin{tabularx}{\textwidth}{ccccX}
    \hline
        Constraint & Mass Regime & Redshift & $\log_{10}\left(\frac{M_{ev}(z)}{M_{\odot}}\right) $ & References\\
        \hline
        Big Bang Nucleosynthesis & $-24.3 < \log_{10}\left(\frac{M}{M_{\odot}}\right) < -19.8$  & $\sim 10^{10}$ & $-25.1$ & \citet{Zeldovich:1977,Carr:2010} \\
        Extragalactic $\gamma$-ray background & $-18.6 < \log_{10}\left(\frac{M}{M_{\odot}}\right) < -16.2$  & $1160$ & $-20.6$ & \citet{Carr:2016_gammaray} \\
         INTEGRAL & $-18.3 \lesssim \log_{10}\left(\frac{M}{M_{\odot}}\right) < -16.1$  & $0$ & $-19.1$ &  \citet{Dasgupta:2019,Laha:2019PhRvL,DeRocco:2019PhRvL} \\
         GRB lensing$^*$ & $-16.3 \lesssim \log_{10}\left(\frac{M}{M_{\odot}}\right) \lesssim -14$  & $0$ & $-19.1$ &  \citet{Marani:1999,Nemiroff:2001,Barnacka:2012,Katz:2018}\\
         White dwarfs$^*$ & $-14.9 \lesssim \log_{10}\left(\frac{M}{M_{\odot}}\right) \lesssim -10.8$  & $0$ & $-19.1$ &  \citet{Graham:2015} \\
         Neutron star capture$^*$ & $-14.9 < \log_{10}\left(\frac{M}{M_{\odot}}\right) < -8.3$  & $0$ & $-19.1$ &  \citet{Capela:2013,Montero-Camacho:2019} \\
        Subaru$^{*}$ & $-11.4 \lesssim \log_{10}\left(\frac{M}{M_{\odot}}\right) \lesssim -5.2$  & $0$ & $-19.1$ &  \citet{Niikura:2019subaru,Smyth:2020} \\
        MACHOS & $-8 \lesssim \log_{10}\left(\frac{M}{M_{\odot}}\right) < 1.8$  & $0$ & $-19.1$ &  \citet{Paczynski1986} \\
        EROS & $-7.2 \lesssim \log_{10}\left(\frac{M} {M_{\odot}}\right) < 0.7$  & $0$ & $-19.1$ &  \citet{Hamadache:2006,Tisserand:2007} \\
        OGLE & $-6.7 \lesssim \log_{10}\left(\frac{M}{M_{\odot}}\right) < -0.5$  & $0$ & $-19.1$ &  \citet{Ogle:2011} \\
         Accretion of PBHs$^*$ & $0 < \log_{10}\left(\frac{M}{M_{\odot}}\right) < 4$  & $450$ & $-20.4$ &  \citet{Poulin:2017,Serpico:2020,Carr:2020arXiv200212778C} \\
        Gravitational waves$^*$ & $1 < \log_{10}\left(\frac{M}{M_{\odot}}\right) < 3$  & $0$ & $-19.1$ &  \citet{Abbott:2018,Wang:2018,Boehm:2020, Raidal:2019JCAP, Wong:2021PhRvD} \\
        Large scale structure & $2 < \log_{10}\left(\frac{M}{M_{\odot}}\right) < 14$  & $0$ & $-19.1$ &  \citet{Carr:2020arXiv200212778C} \\
         Lensing of radio sources & $5 < \log_{10}\left(\frac{M}{M_{\odot}}\right) < 9$  & $0$ & $-19.1$ &  \citet{Wilkinson:2001PhRvL} \\
         Dynamical friction & $4 \lesssim \log_{10}\left(\frac{M}{M_{\odot}}\right) < 13$  & $0$ & $-19.1$ &  \citet{Carr:2020arXiv200212778C} \\
         Wide binaries & $2.7 \lesssim \log_{10}\left(\frac{M}{M_{\odot}}\right) < 8$  & $0$ & $-19.1$ &  \citet{Chaname:2004ApJ,Quinn:2009MNRAS} \\
        X-ray binaries & $0.8 \lesssim \log_{10}\left(\frac{M}{M_{\odot}}\right) \lesssim 7.3$  & $0$ & $-19.1$ &  \citet{Inoue:2017,Carr:2020arXiv200212778C} \\
        Globular cluster disruption & $4.5 < \log_{10}\left(\frac{M}{M_{\odot}}\right) < 11$  & $0$ & $-19.1$ &  \citet{Carr:1999ApJ_dynamical} \\
       Galaxy disruption & $9.8 < \log_{10}\left(\frac{M}{M_{\odot}}\right) < 13$  & $1$ & $-19.2$ &  \citet{Carr:1999ApJ_dynamical} \\
       Disk heating & $6.5 < \log_{10}\left(\frac{M}{M_{\odot}}\right) < 12.5$  & $1$ & $-19.2$ &  \citet{Carr:1999ApJ_dynamical,Carr:2020arXiv200212778C} \\
       CMB dipole & $16.8 \lesssim \log_{10}\left(\frac{M}{M_{\odot}}\right) \lesssim 22$  & $0$ & $-19.1$ &  \citet{Carr:2020arXiv200212778C} \\
    \hline
    \end{tabularx}
    \label{tab:constraints}
\end{table*}

\subsubsection{Dynamical friction}
PBHs could be dragged into the centre of the Milky Way due to dynamical friction of halo objects and stars. This possibility leads to constraints on $f$ in the range of masses between $\sim 10^{4} M_{\odot}$ and $\sim 10^{13} M_{\odot}$. To use our method to translate monochromatic constraints to extended ones, we extract the data points from the $f$ curve presented by \citet{Carr:2020arXiv200212778C}.

\subsubsection{Accretion by PBHs}

The accretion of matter onto PBHs involves different effects that we can potentially observe. Even if there are numerous constraints related to this process, we consider here the constraints on PBHs with masses between $\sim 10^0 $ and $ 10^4\, M_\odot$ , presented by \citet{Serpico:2020}. In particular, they study the effects of disk-like or spherical accretion on the CMB anisotropies. In this work, we adopt the accretion scenario without a DM halo\footnote{This is because we start with the assumption that DM is composed by PBHs and then, we study the validity of this assumption by computing $f$.} \citep[explained in detail by][]{Poulin:2017}. The relevant redshift for this process is considered as $z \sim 450$.

In general, constraints related to the accretion by PBHs depend on numerous assumptions and physical parameters. Therefore these must be considered with care \citep[see][]{Carr:2020arXiv200212778C}.

\subsubsection{Large scale structure}

Massive PBHs have the peculiarity that they can seed the large scale structure of the Universe \citep{Carr:2018}. To take this into account we consider the estimations for the constraint on $f$ for this effect, given by \citet{Carr:2020arXiv200212778C} who include constraints for PBH with masses in the range $\sim 10^2 - 10^{14} \, M_\odot$. The relevant redshift for this constraint is $z = 0$.

\subsubsection{Gravitational waves}

Gravitational waves are produced by the coalescence of black holes, which may be primordial in origin. They could also be produced during the formation of PBHs.  Several authors studied the merger rates of PBHs in order to predict GW signals due to these mergers and compared them to the observations \citep[see][for example]{Sasaki:2016,Eroshenko:2018}. \citet{Ali-Haimoud:2017} estimated the merger rate of PBH binaries in order to compute the maximum fraction $f$, obtaining potential constraints on PBHs with masses in the range $\sim 10 - 100 M_{\odot}$. Later, the LIGO/Virgo collaboration used the non-detection of GW events to put constraints on sub-solar mass PBHs \citep{Abbott:2018}.

Additionally, the superposition of GW from independent sources produces a background signal known as stochastic gravitational wave background (SGWB). \citet{Wang:2018} used this effect to compute constraints on $f$ for  PBHs of$\sim 1 - 100 M_\odot$, using the first Advanced LIGO observation run. \JSH{Recently, \citet{Raidal:2019JCAP} provide updated constraints in the mass range $\sim 10^{-1}-10^{3}$ (assuming a log-normal mass function) from the observed merger rate of ten events by LIGO non-observations and also bounds from the stochastic GW background by comparing with the projected final sensitivity of LIGO. \citet{Wong:2021PhRvD} present constraints on $f$ estimated from the third observing run of the LIGO-Virgo Collaboration and from the NANOGrav experiment $11$-yr data. We have also considered these bounds in our analysis.}

Even if these effects impose stringent constraints on $f$, it has been recently pointed out that a more detailed analysis to compute the merger rate of PBH binaries is needed \citep[see][]{Boehm:2020}. This result suggests that the constraints related to GW must be disputed if these are calculated by estimating a merger rate for PBH binaries.
\JSH{Nevertheless, several authors \citep{DeLuca:2020JCAP,Hutsi:2021} claimed that such analysis on the growing PBH mass in an expanding universe should be reexamined.}

\subsubsection{Cosmic microwave background dipole}
Under the assumption that there are supermassive PBHs in the intergalactic medium, they can induce peculiar velocities on galaxies due to gravitational interaction. The peculiar velocity of the Milky Way can be measured from the cosmic microwave background (CMB) dipole and used to constrain the fraction of this population of PBHs \citep{Carr:2020arXiv200212778C}. The resulting constraint on $f$ gives

\begin{equation}
f_{CMB}(M)=\left(\frac{M}{5\times 10^{15} M\odot}\right)^{-1/2} \left(\frac{t_{0}}{10^{10} \mathrm{yr}}\right)^{-3/2} \Omega_{m0}^{-0.9}h^{-2},
\end{equation}

\noindent
where $t_{0}$ is the age of the Universe, $\Omega_{m0}$ is the matter density parameter and, $h$ is the dimensionless normalised Hubble constant. Even though the CMB radiation originates at $z \sim 1100$, this effect is measured locally, implying that the relevant redshift is $z=0$. This effect gives constraints on PBHs with masses between $\sim 7\times10^{16} M_{\odot}$ and $1\times10^{22}M_{\odot}$. 

Although we have considered this limit, we will explore the $M_{*}$ parameter up to $10^{15} M_{\odot}$, meaning that it is very unlikely to find PBHs within the mass regime of this process and hence, we do not expect to obtain an $f(M)$ bound from this constraint.

In Table \ref{tab:constraints} we summarise these physical effects and the corresponding PBH masses that each one constrains, along with the relevant redshift, the mass of the PBH that evaporates at that time $M_{ev}(z)$ and references to the full details of these effects. 

\JSH{ As mentioned before, \citet{Carr:2017} and \citet{Bellomo_2018} presented different methods to translate constraints on $f$ from monochromatic mass functions to extended ones. The method by \citet[CM17,][]{Carr:2017} computes the $f$ of an extended mass distribution by integrating the quotient between the mass distribution of $log(M)$ and the maximum allowed fraction for a monochromatic function. The method by \citet[BM18,][]{Bellomo_2018} consist in estimating $f$ from an effective mass, $M_{eff}$, associated to a monochromatic mass function. This $M_{eff}$ is calculated by integrating the mass function normalised to unity weighted by a function $g(M)$ encoding the physical processes of the observable. To compare these methods with the one presented in this work use the observations of gravitational lensing by MACHOS (see \S \ref{subsec:lensing_constraints}). The resulting $f$ values for selected $(n_b,M_{*})$ values are shown in Table \ref{tab:values_f_comparison}. It is worthy to mention that when the CM17 and BM8 methods are applied to our extended mass functions, the $f$ is calculated only in the range of masses of the chosen observables. We found that the $f$ values obtained from the MACHOs constraints are consistent among the three different methods. It is worth noting that the method presented in this work corrects the $f$ value taking into account  PBHs outside the limits of the observable universe as a function of redshift, as well as the evolution of the mass function from the redshift $z$ to the current epoch through $M_{ev}(z)$ and $M_{1ph}(z)$. In addition, we found that $f$ from BM18 is almost independent from the $M_{*}$ value. Moreover, if $f$ is calculated considering the minimum and maximum masses of the full extended mass distribution, this method could result in $M_{eff}$ values outside the range of masses of the observable, and therefore, other considerations must be taken into account.
\begin{table}
\caption{Comparison of the allowed fractions of PBHs as DM obtained from different methods to translate results for monochromatic constraints to extended mass functions for the set of parameters given in Table \ref{tab:values_f1} using the gravitational lensing by MACHOS.}
    \centering
    \resizebox{0.48\textwidth}{!}{%
    \begin{tabular}{ccccc}
    \hline
        $n_{b}$ & $M_{*}[M\odot]$ & \multicolumn{3}{c}{$f_{\rm{MACHOs}}$} \\
    \hline
    & &This work& \citet{Bellomo_2018} & \citet{Carr:2017}\\
    \cline{3-5}\\
        \multicolumn{5}{c}{\textbf{HC}}\\
        $3.0$  & $1.39\times10^{2}$ &$1.0$ &$0.68$ & $1.0$\\
     $3.8$  & $9.1\times 10^{-8}$ & $1.0$&$0.83$ & $1.0 $ \\ \\
        \multicolumn{5}{c}{\textbf{FCT}}\\
        $1.5$ & $1.39 \times10^{2}$ &$1.0$& $0.31$ & $1.0$ \\
    $3.5$ & $1.39 \times10^{2}$ & $1.0$ &$0.10$ & $0.86$\\
    $2.0$ & $9.1 \times10^{-8}$ & $1.0$ &$0.81$ & $1.0 $\\
    \hline
    \end{tabular}
    }
    \label{tab:values_f_comparison}
\end{table}
}

\section{Combined constraints on extended PBH mass functions}
\label{sec:results}
To investigate whether PBHs under the HC and FCT
formation scenarios can constitute all the dark matter in the Universe we confront different realisations for the mass functions with observational constraints at different mass regimes, mentioned in the previous section.  As mentioned above, we assume fiducial values $n_{s}=0.9649$, $k_{piv}=10\,\mathrm{Mpc^{-1}}$, and $f_{m}=1$ in the HC (Eq. \ref{eq:dndmHC}) and FCT (Eq. \ref{eq:dndmFCT}) mass distributions, leaving as free parameters the blue index $n_{b}$ and $M_{*}$.\footnote{It is worth to note that our first choice on $f_{m}$ implies that only a fraction of regions with a \JS{linear} overdensity higher than $\delta_{c}$ will \JS{be associated to the physical formation of} a PBH.} Each realisation of the mass function will have a different pair ($n_{b}$, $M_{*}$) spanning the intervals  $n_{b} \in [1.1,4.0]$ and $M_{*}\in[10^{-30}M_{\odot},10^{15} M_{\odot}]$ in a grid of 50 points for each parameter.  

For all the realisations, we compute the constraint on $f_{\text{SMBH}}$ (Eq. \ref{eq:fSMBH}), as explained above. Additionally, we employ the method described in Section \ref{sec: MMD to EMD} for obtaining the corresponding constraint on $f_i$ for each process described in the same section. 

\begin{figure*}
    \centering
    \includegraphics[width=13cm, height=10cm]{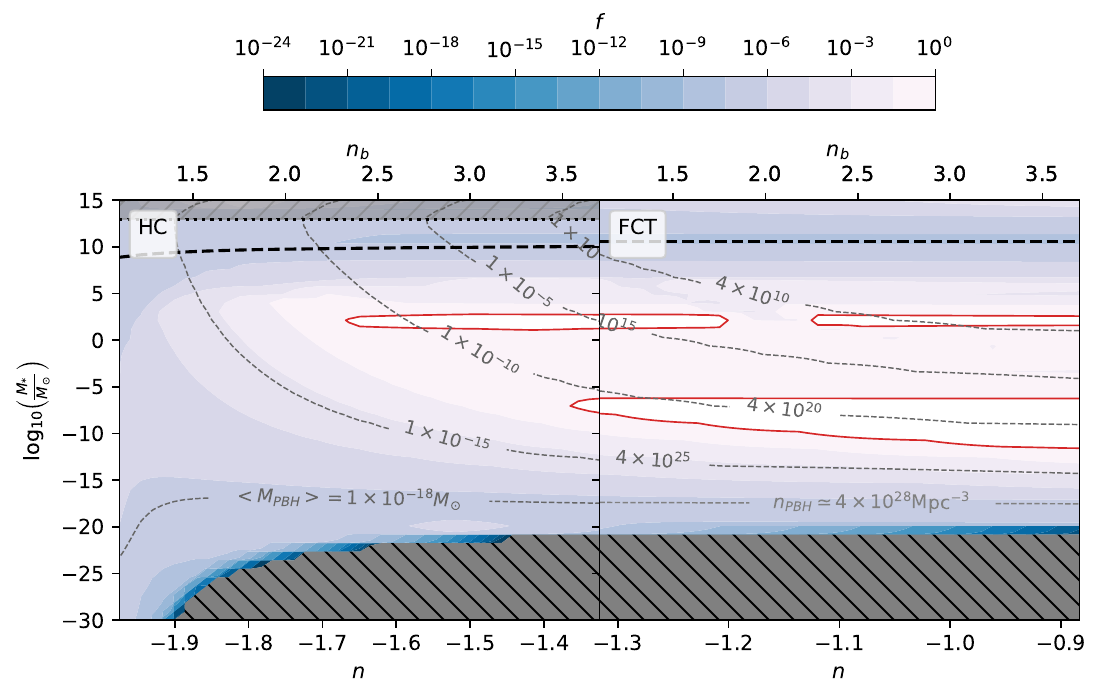}
    \\
    \includegraphics[width=13cm]{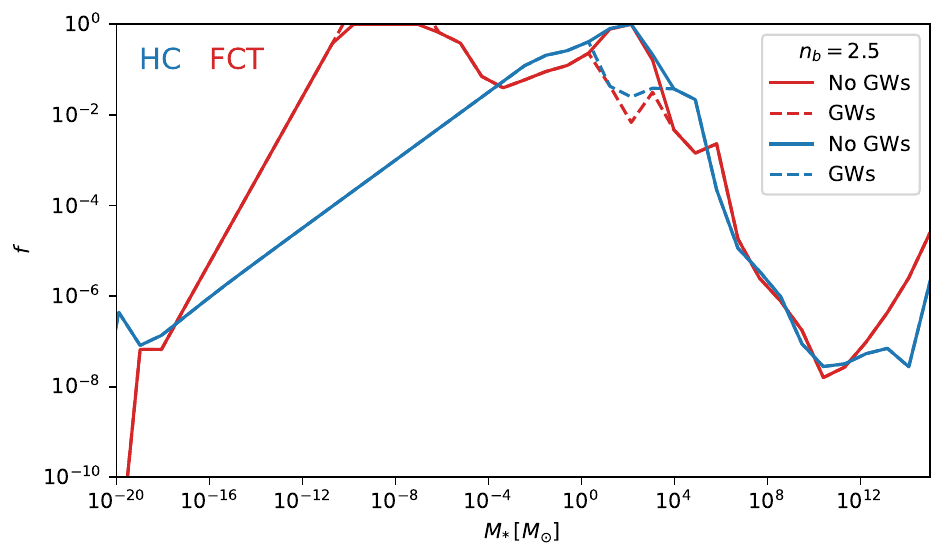}
    \caption{Top panel: level contours for the fraction of PBHs as DM in HC (left panel) and FCT (right panel) for different values of the slope of the mass function and the characteristic mass $M_{*}$. The upper secondary axis gives the values for the blue index $n_{b}$. The red lines correspond to $f=1$ and the allowed areas are represented in white. The grey dashed contours in the HC and FCT panels  represent the values of the average mass $\left\langle M\right\rangle_{PBH}$  (Eq. \ref{eq:average_mass}) and the number density $n_{PBH}$ (Eq. \ref{eq:NPBH}) of the PBH distributions, respectively. The hatched region for $M^{*} \lesssim 10 ^{-21}\, M_\odot$ is excluded since those mass functions predict that all of the PBHs have evaporated by the present time. In the HC scenario, the top dotted lines correspond to $M_{piv}=8.9\times10^{12} M_{\odot}$ when $f_{m}=1$. The hatched region for $M_{*}>M_{piv}$ shows there are no HC mass functions defined there. The black dashed line in both scenarios show $M_{piv}$ for $f_m=\beta$. \JSH{Bottom panel: fraction of PBHs as DM $f$ for fixed $n_b=2.5$ as a function of $M_*$. Red lines correspond to the FCT scenario and blue lines to the HC scenario. Dashed lines show the resulting $f$ including constraints from GW.}}
    \label{fig:constraints}
\end{figure*}

Figure \ref{fig:constraints} (top panel) shows the resulting level contours for the maximum allowed fraction $f$ of DM in PBHs obtained by combining the undisputed monochromatic constraints (see Table \ref{tab:constraints}) according to Eq. \ref{eq: minimum f}, together with the constraint provided by SMBHs. The colours correspond to contours on $f$ for values between $10^{-24}$ and $1$. We present this as a function of the slope $n$ of the PBH mass functions (bottom axis), $n_b$ (top axis) and $M_{*}$. The left and right panels show the HC and FCT scenarios, respectively.
In the HC panel, the grey dashed contours represent the values of the average mass $\left\langle M\right\rangle_{PBH}$  (Eq. \ref{eq:average_mass}) of the PBH distributions with values from $10^{-18} M_{\odot}$ to $1 M_{\odot}$ from the bottom to the top, respectively. In the FCT panel, the grey dashed contours represent the values of the number density $n_{PBH}$ (Eq. \ref{eq:NPBH}) with values from $\simeq 4 \times 10^{28} \text{Mpc}^{-3}$ to $\simeq 4 \times 10^{10} \text{Mpc}^{-3}$ from  bottom to  top, respectively.
In both panels, the hatched region at low characteristic masses ($M_{*} \lesssim 10
^{-21}\, M_\odot$), is excluded since in this range of parameters all PBHs have evaporated by the present time and hence, cannot account for the DM we see today. As can be seen, a large fraction of the parameter space is restricted to $f<1$. However, the red contours that correspond to $f=1$ enclose  regions (white) where it is possible to have the DM composed entirely by PBHs, i.e., $f \geq 1$. The are two allowed regions in the HC scenario, one of them roughly at $M_{*}\sim 10^{2} M_{\odot}$, with $n_{b}$ from $\sim2.3$ to $4$. The other region corresponds to $M_{*}\sim10^{-7} M_{\odot}$ and $n_{b}>3.6$. In the FCT scenario, there are three allowed regions where $f=1$. The first two regions are located at $M_{*}\sim 10^{2} M_{\odot}$ and $n_{b}$ in the ranges $[1.1,1.7]$ and $[2.2,4.0]$ respectively. The third region spans all the $n_{b}$ range for $M_{*}\sim10^{-7} M_{\odot}$ including even more values of $M_{*}$ as $n_{b}$ increases. Table \ref{tab:values_f1} gives  representative $(n_{b}, M_{*})$ values in such regions. 

An interesting feature is that the level contours exhibit a continuity in the mass function slope from $n \sim -1.9$ to $-0.9$. In the HC scenario, the top dotted lines correspond to a pivot mass $M_{piv}=8.9\times10^{12} M_{\odot}$ when $f_{m}=1$. The hatched region for $M_{*}>M_{piv}$ shows there are no HC mass functions defined there.  

Regarding $f_{m}$,  we also perform an iterative procedure to estimate \JS{$f_m = \beta$ such that} all \JS{linear} overdensities with $\delta > \delta_{c}$ \JS{are associated with the formation of} a PBH. \JS{Applying the procedure explained in Section \ref{sec:app_fm}, in FCT} we obtain $\beta\sim 5.8 \times 10^{-23}$\JS{, for all values of $n_b$ and $M_*$,} resulting in $M_{piv} \sim 3.45 \times 10^{10} M_{\odot}$ which is shown \NP{as} the black dashed line in the FCT panel. \JS{In the HC scenario, this is slightly more complicated due to the strong dependence of $\beta$ on $M_*$. The approach for this scenario, was to find the maximum $M_*$ value such that $M_*<M_{piv}$. The resulting}  maximum effective $\log_{10}(M_{*}/M_{\odot})\sim 8.8-9.8$ in the $n_{b}$ range implying $M_{piv}\sim 7.78\times 10^{8} M_{\odot}-1.2 \times 10^{10} M_{\odot}$. The black dashed line in the HC panel shows these values of maximum $M_{*}$. 
 
\JSH{The bottom panel of Figure \ref{fig:constraints} shows the combined constraints for $f$ for a fixed value of $n_b = 2.5$ as a function of $M_*$ for both, the FCT (red) and HC (blue) scenarios. In this figure, we also show the value of $f$ including (solid lines) and excluding (dashed lines) the constraints from GW, where the two cases differ mostly in the region around $M_* \sim (10^1 - 10^4) M_\odot$.  The inclusion of the GW constrains exclude the possibility that $f=1$ in this range of $M_*$. We emphasise that the relation between $f$ and $M_*$ is different from the usual $f(M)$ for monochromatic PBH distributions because, in this case, we consider an extended mass distribution; i.e., choosing a different $n_b$ value will change drastically the results for $f(M_*)$.}
 
We emphasise that these results for $f$ are obtained discarding the disputed constraints. If one includes them without any additional considerations, PBHs as the sole component for DM get completely ruled out on the entire parameter space for both scenarios. In particular the NS constraint erases most of the allowed region around $M_{*} \sim 10^{-7}\, M_{\odot}$ for the FCT scenario and the Subaru constraint completely eliminates this region. Also, the constraints from the \JSH{non-observation of SGWB \citet{Wong:2021PhRvD}, mergers \citet{Raidal:2019JCAP} and accretion by PBHs \citet{Poulin:2017, Serpico:2020} eliminate the region around $M_{*} \sim 10
^{2}\, M_{\odot}$ in both scenarios. Particularly, the inclusion of the GW constraints for the parameters of Table \ref{tab:values_f1} changes the $f$ values from $1$ to $\sim 10^{-2}$ (similar values are obtained including the accretion constraints)}. Therefore, understanding the physics of these constraints becomes crucial if we want to rule out PBHs as a DM candidate.
\JSH{Notice that when the monochromatic constraints are combined, PBH with masses $10^{-16}-10^{-11} M_{\odot}$ can comprise an important fraction (or the total) of the content of dark matter in the Universe \citep{Carr:2020arXiv200212778C, Carr:2020b}. In the case of extended distributions, $M_*$ values in this range are restricted mainly by
the galactic centre $\gamma$-ray (INTEGRAL) and extragalactic $\gamma$-ray background constraints, in the FCT scenario, and these same bounds in combination with those from BBN, in the HC scenario. However, it should be noted that $M_* \neq M_\mathrm{PBH}$ where $M_\mathrm{PBH}$ is the mass of a PBH in a monochromatic population. The fact of $M_*$ being close to the maximum mass of the distribution implies that, for a given value of $M_*$, all the constraints that affect lower masses must be considered. In this particular case, the abundance of PBHs with masses between $(10^{-24} - 10^{-16})M_\odot$ is still high enough to disagree with the observations from the galactic centre $\gamma$-rays, extragalactic $\gamma$-ray background, and BBN for values of $M_*$ between $(10^{-16} - 10^{-11})M_\odot$, thus ruling out that region of the parameter space.}
\subsection{Further analysis in the allowed windows}

To check whether the parameters in the regions that allow all of DM in PBHs make sense physically, we evaluate the ratio $\delta_c/\left<|\delta|^2\right>^{1/2}$ (See Section \ref{sec:app_ratiodeltas}) within them and show the results in Table \ref{tab:values_f1}.  The large values found for this ratio for all allowed windows in both scenarios indicate that the less preferred possibility of having $f_m=1$ can be discarded. Considering  $f_m = \beta$ (i.e., that all regions will collapse into a PBH but only with a fraction $f_m$ of its energy density) 
we see that for the HC scenario we obtain $\delta_c/\left<|\delta|^2\right>^{1/2}\sim 1$, at least for the window centered at $M^*\sim 100M_\odot$. The second \NP{HC} window \NP{with $M^*\sim 10^{-8} M_\odot$} is rejected since, even when considering $f_m=\beta$ we obtain $\delta_c \gg \left<|\delta|^2\right>^{1/2}$. 
\JSH{In the FCT windows that allow all DM in PBHs, we find a similar result for $f_m=\beta$, namely, $\delta_c \sim \left<|\delta|^2\right>^{1/2}$ in the $M_*\sim 100 M_\odot$ window and  $\delta_c \gg \left<|\delta|^2\right>^{1/2}$ in the second window. 
We explored changing the scale factor $a_{fct}$ by several orders of magnitude and found that the resulting values of $\delta_c/\left<|\delta|^2\right>^{1/2}$ do not change significantly and remain within roughly the same order of magnitude.
}

We point out that in the evaluation of these quantities, the contribution of neutrinos and other relativistic species is taken into account, since it may affect the results for $\delta_c/\left<|\delta|^2\right>^{1/2}$ and $\beta$.


\begin{table}
\caption{Characteristic mass ($M_{*}$)\JS{, blue index ($n_{b}$), and ratio between the corresponding critical density contrast over the typical overdensity $\left(\delta_c/\left<|\delta|^2\right>^{1/2}\right)$} 
where all DM can be composed by PBHs, i.e. $f=1$, \JS{ for $f_m = 1$ and  $f_m=\beta$.}}
    \centering
    \begin{tabular}{cccc}
    \hline
        $n_{b}$ & $M_{*}[M\odot]$ & \multicolumn{2}{c}{$\delta_{c}\JS{/\left<|\delta|^2\right>^{1/2}}$} \\
    \hline
    & & $f_m = 1$ & $f_m=\beta$\\
    \cline{3-4}\\
        \multicolumn{4}{c}{\textbf{HC}}\\
        $3.0$  & $1.39\times10^{2}$ & $1.28\times10^{5}$ & $1.13$\\
     $3.8$  & $9.1\times 10^{-8}$ & $2.63\times 10^{13}$ & $1.29\times 10^{5}$ \\ \\
        \multicolumn{4}{c}{\textbf{FCT}}\\
        $1.5$ & $1.39 \times10^{2}$ & $5.27\times 10^{14}$ & \JSH{$1.07$} \\
    $3.5$ & $1.39 \times10^{2}$ & $4.50\times10^{15}$ & \JSH{$9.15$}\\
    $2.0$ & $9.1 \times10^{-8}$ & $3.65\times10^{20}$ & \JSH{$7.40\times 10^{5}$}\\
    \hline
    \end{tabular}
    \label{tab:values_f1}
\end{table}

\section{Conclusions} \label{sec:conclusions}

In this paper we \NP{used a modified Press-Schechter (PS) formalism to} investigate the possibility that primordial black holes make up a fraction of the dark matter in the Universe. \JS{We modified the standard Press-Schechter formalism used to infer the abundance of DM haloes so that it can be applied to PBH formation since, in this case, there is no simple, known relation between the linear overdensity and the physics of PBH formation in an expanding background. 
This results in extra parameters for PS the first of which is the fraction of the overdensity to undergo collapse, $f_m$.} \NP{This parameter allows the formalism to use all the information from linear theory when it takes a value corresponding to the ratio of dark-matter to total energy densities at the median formation time.}
\NP{A second modification is that we }\JS{also allow the relevant amplitudes of linear fluctuations to be measured either at a fixed conformal time or at horizon crossing. Following the modified Press-Schechter formalism, under both formation timings, we considered two primordial power spectra, the standard one and a broken power law to obtain the mass functions in terms of two additional free parameters, the blue index $n_b$ and the pivot scale $k_{piv}$.  We also use the standard  \JS{linear  density contrast} threshold for collapse $\delta_c$.  \NP{This parameter is particularly meaningful since for linear theory to remain informative about the non-linear collapse, its value should be of the same order as the typical energy density fluctuations at the time of collapse.}  All these parameters are encoded in the characteristic mass, $M_{*}$, and mass function slope $n$.} 

To restrict these parameters, we introduce a new constraint for extended PBH mass distribution employing the SMBH mass function. This constraint arises from requiring that the abundance of massive PBHs should in no case exceed the abundance of SMBHs. This naturally imposes strong constraints for the high characteristic masses of $M_{*} > 10^{10} M_{\odot}$. In addition, we used several observational constraints at  different mass ranges coming from monochromatic PBH mass functions. 
We introduce a new approach to apply these monochromatic constraints to extended mass distributions through an output function $g(M)$ encoding the underlying process for each observable such that the resulting allowed fraction is the inverse of the output i.e. $f \sim 1/g(M)$ (see Section \ref{sec: MMD to EMD}). Moreover, we consider two new corrections to the PBH mass fraction $f$, which take into account the fraction of the mass that is constrained by any particular physical process ($C_{M}$) and the redshift evolution of the mass function ($C_{z}$). We only consider PBHs such that they exist (e.g., not evaporated at the relevant redshift) within the causal volume. By using this formalism, we obtain that the bounds obtained in monochromatic mass function are weaker when they are translated to extended mass distributions. 
To obtain the final (and most restrictive) constraints on the free parameters $M_{*}$ and $n_{b}$ we compute the maximum allowed fraction as the minimum one out of all the undisputed constraints for that particular choice of mass function parameters. 

For both, HC and FCT  mass functions, we obtain that there are potential regions where all DM can be made of PBHs. In the HC scenario, these regions roughly correspond to a characteristic mass $M_{*}$ of $\sim 10^{2} M_{\odot}$ with $n_b$ from $\sim 2.3$ to $4$ and an $M_{*}$ of $\sim10^{-7} M_{\odot}$ with $n_b >3.6$.  
In FCT the fraction $f=1$ is allowed when $M_{*}\sim 10^{2} M_{\odot}$ for $1.1<n_{b}<1.7$, or $2.2< n_{b}<4.0$, and $M_{*}\sim10^{-7} M_{\odot}$ for all explored values of $n_{b}$.

We also verify if the linear perturbations are related to the physics of collapse, within the allowed windows, by evaluating $\delta_c/\left<|\delta|^2\right>^{1/2}$. We do this calculation considering $f_m = 1$ and $f_m = \beta$, where the latter is the preferred option as it ensures a strong relation between PBH formation and the linear perturbations. 
As indicated in Table \ref{tab:values_f1}, 
the \JSH{windows} centered at $M_*\sim100 M_\odot$ \JSH{are} the only ones to satisfy the condition of $\delta_c/\left<|\delta|^2\right>^{1/2}\sim1$ for $f_m=\beta$.

We emphasise that the \NP{allowed, and physically sensible, HC} region for $M_{*}\sim 10^2$ for a wide range of $n_{b}$ is interesting as it is of the order of black hole masses ($\sim 30 M_{\odot}$) measured by LIGO from the gravitational waves of a binary black hole merger \citep{Abbott:2016blz,Jedamzik:2020}. It is worth to note, that more stringent bounds on the characteristic mass $M_{*}$ and $n_{b}$ could could be obtained if more observational constraints are considered, for example, if we consider all the disputed bounds, indicated by $^*$ in Table \ref{tab:constraints}, we completely rule out that DM is fully composed by PBHs in the scenarios studied in this work. 

In summary, we have revisited the PS formalism to construct extended PBH mass distributions under two formation scenarios (i.e., FCT and HC). 
In constructing the mass function using the PS formalism, we have considered a broken power-law primordial power spectrum with a blue index for small scales.
When neglecting the disputed constraints, we found regions for the mass function parameters which allow all the DM in the Universe to be made of PBHs. We encourage further investigation of these models to elucidate the true nature of dark matter.

\section*{Acknowledgements}

We thank Nicola Amorisco, Carlton Baugh, Julio Chanam\'e, Carlos Frenk, Baojiu Li, Jorge Nore\~na, Loreto Osorio,  Marco San Mart\'in, Dom\'enico Sapone, and Jakub Scholtz for helpful discussions.  We thank the anonymous Referee for their helpful comments.  This project has received funding from the European Union's Horizon 2020 Research and Innovation Programme under the Marie Sk\l{}odowska-Curie grant agreement No 734374.
NP wants to thank the hospitality of the Institute for Advanced Studies at Durham University (UK) and its Fellows programme, during which part of this work was carried out.
NP, JM and JS acknowledge support from CONICYT project Basal AFB-170002. NP and JS were supported by Fondecyt Regular 1191813.
The work of IJA is funded by ANID, REC Convocatoria Nacional Subvenci\'on a Instalaci\'on en la Academia Convocatoria A\~no 2020, Folio PAI77200097.
The calculations presented in this work were performed on the Geryon computer at the Center for Astro-Engineering UC, part of the BASAL PFB-06 and AFB-170002, which received additional funding from QUIMAL 130008 and Fondequip AIC-57 for upgrades.

\section*{DATA AVAILABILITY}

No new data were generated or analysed in support of this research.





\bibliographystyle{mnras}
\bibliography{references} 



\appendix

\section{Black Hole Evaporation}\label{sec:AppendixBHevap}

The power radiated by an object with temperature $T$ is given by

\begin{equation}
    P = A\epsilon \sigma T^4,
    \label{eq: Power}
\end{equation}

where $A$ is the radiating surface area of the object, $\epsilon$ is the emissivity and $\sigma$ is the Boltzmann constant, defined as

\begin{equation}
    \sigma = \frac{\pi^2 k_B^4}{60 \hbar^3 c^2}.
    \label{eq: Boltzmann constant}
\end{equation}

For a black hole, we can consider the simple scenario of a Schwarzschild Black Hole. Then, its temperature is directly related to the mass of the BH $M$ by

\begin{equation}
    T(M) = \frac{\hbar c^3}{8\pi G k_B M}.
    \label{eq: Hawking Temp}
\end{equation}

In this situation, the radiating surface area of the BH is given by the surface area of a sphere with radius $r$ equal to the Schwarzschild radius $r_s$ of the black hole. This is

\begin{equation}
    A = 4\pi r_s^2 = \frac{16\pi G^2 M^4}{c^4},
    \label{eq: surface area of a black hole}
\end{equation}

where the Schwarzschild radius is given by $r_s = 2GM/c^2$. Considering that the black hole radiates as a perfect black body ($\epsilon = 1$), the Power radiated by a black hole becomes

\begin{equation}
    P = \frac{\hbar c^6}{15360 \pi G^2 M^2}.
    \label{eq: Power BH}
\end{equation}

The power radiated by the black hole is nothing more than the rate of energy loss by the BH, and this, in turn, corresponds to the rate of mass loss for the BH

\begin{equation}
    P = -\frac{dE}{dt} = -c^2 \frac{dM}{dt}.
\end{equation}
Then, we can write

\begin{equation}
    -\frac{dM}{dt} = \frac{\hbar c^4}{15360 \pi G^2 M^2}
\end{equation}
and integrating this expression, we have that 

\begin{equation}
    t(M) = \frac{5120 \pi G}{\hbar c^4} \left(M_0^3 - M^3\right).
    \label{eq: time evaporation}
\end{equation}

$t(M)$ indicates the time needed for a black hole to go from its initial mass $M_0$ to the mass $M$. Evaluating this expression for $M=0$, the evaporation lifetime $\tau_{ev}$ (Eq. \eqref{eq: evaporation lifetime}) is recovered. 

We can evaluate this expression to find how much time it takes to a black hole to evaporate half of its mass.

\begin{equation}
    t\left(\frac{M_0}{2}\right)= \frac{5120 \pi G}{\hbar c^4} \left(M_0^3 - \frac{M_0^3}{8}\right),
\end{equation}

where if we use the expression for $\tau_{ev}$, we have that

\begin{equation}
     t\left(\frac{M_0}{2}\right)= \frac{7}{8}\tau_{ev}.
\end{equation}

Then, we can conclude that, independently of the initial mass $M_0$ of the BH, the time needed to evaporate half of its mass will be $7/8$ of its complete lifetime. Therefore, a BH will evaporate its second half of its mass within the last eighth of its lifetime.

\section{Further mathematical details}
\label{sec:appA}
Here we give further details of all our analytical expressions for the mass functions.

\subsection{Fixed conformal time}\label{sec: appendix FCT}

\subsubsection{Standard Power Spectrum}
The variance of the density field \eqref{eq:sigmasquared} for the power law spectrum \eqref{eq:pkplaw} results in
\begin{eqnarray}
\sigma^{2}(k)&=&4\pi D^2(a) \int\limits_{0}^{k_R} P(k)k^{2} \mathrm{d}k=  \frac{4\pi\,D^2(a)\,A_{s}}{k_{0}^{n_{s}}}\int\limits_{0}^{k_{R}} k^{n_s+2} dk\nonumber\\
&=& \frac{4\pi\,D^2(a) A_{s}}{k_{0}^{n_{s}}} \frac{k_{R}^{n_s+3}}{n_s+3},
\label{eq:sigma_plaw_fct}
\end{eqnarray}
where $D^2(a)=a^4_{fct}$ since $a_{fct}$ lies within radiation domination. 

By substituting the relation of $k_{R}$ in terms of the mass given by \eqref{eq:k_R__fct},
Eq. \eqref{eq:sigma_plaw_fct} can be rewritten  as

\begin{equation}
\sigma^{2}(M)=\frac{4\pi\, a^4_{fct}\,A_{s}}{k_{0}^{n_{s}}} 
\frac{ C_{fct}^{n_s+3} \JS{f_m^{\frac{n_s+3}{3}}}}{(n_s+3)} M^{-\frac{(n_s+3)}{3}}.
\end{equation}

In addition, the $\nu$ parameter \eqref{eq:nu} reads as
\begin{equation}
    \nu= \frac{\delta_{c}}{\sigma (M)}=\JS{\frac{\delta_{c}}{f_m^{\frac{n_s+3}{6}}}}\sqrt{\frac{ (n_s+3)}{4\pi\, a^4_{fct} (A_{s}/k_0^{n_s}) C_{fct}^{n_s+3}}} M^{\frac{n_s+3}{6}}.
    \label{eq:nuM}
\end{equation}

We define a characteristic mass, $M_{*}$, such that $\nu(M_{*})=1$, given by

\begin{equation}
\JS{\frac{\delta_{c}}{f_m^{\frac{n_s+3}{6}}}}\sqrt{\frac{ (n_s+3)}{4\pi\, a^4_{fct} (A_{s}/k_0^{n_s}) C_{fct}^{n_s+3}}} M_{*}^{\frac{n_s+3}{6}}=1.
\end{equation}

\noindent
Then, the characteristic mass results in
\begin{equation}
    M_* = \left(\sqrt{\frac{4\pi\, a^4_{fct}\,(A_{s}/k_0^{n_s}) C_{fct}^{n_s + 3}}{(n_s +3)}}\right)^{\frac{6}{(n_s +3)}}\JS{\frac{f_m}{\delta_c^{\frac{6}{(n_s +3)}}}}.
    \label{eq: M* plaw fct}
\end{equation}
The threshold density contrast for PBH formation can be expressed as
\begin{equation}
 \delta_{c}=\sqrt{\frac{4\pi\, a^4_{fct}\,(A_{s}/k_0^{n_s}) C_{fct}^{n_s + 3}}{(n_s +3)}} \JS{\left(\frac{f_m}{M_{*}}\right)^{\frac{n_s+3}{6}}}.
\end{equation}
We can rewrite the Eq. \eqref{eq:nuM} in terms of $M_{*}$ as $\nu=\left(\frac{M}{M_{*}}\right)^{\frac{n_s+3}{6}}$ and its derivative with respect to $M$ results as

\begin{equation}
\frac{d\nu}{dM}=\frac{ns+3}{6}\frac{1}{M_{*}}\left(\frac{M}{M_{*}}\right)^{\frac{ns-3}{6}},
\label{eq:dnudm plaw fct}
\end{equation}


\begin{eqnarray}
\left(\frac{dn}{dM}\right)_{\text{fct}}^{\text{std}}&=&\frac{ \rho_{DM}(a)}{\sqrt{2\pi}}\frac{ns+3}{3}\frac{1}{M^{2}}\left(\frac{M}{M_{*}}\right)^{\frac{ns+3}{6}}  \nonumber\\
&&\times \exp\left[-\frac{1}{2}\left(\frac{M}{M_{*}}\right)^{\frac{ns+3}{3}}\right].\nonumber\\
\end{eqnarray}

\subsubsection{Broken Power Spectrum}

The variance of the density field $\sigma^2(M)$, in this scenario is obtained substituting Eq. \eqref{eq:pkbroken} into the definition in Eq. \eqref{eq:sigmasquared}. This reads as

\begin{equation}
\sigma^{2}(M)=
\frac{4\pi  D^2(a) A_{s}}{k_{0}^{n_s}}\left[ \int\limits_{0}^{k_{piv}} k^{n_{s}+2} dk+
  \int\limits_{k_{piv}}^{k_{R}} k_{piv}^{n_{s}-n_{b}}k^{n_{b}+2} dk\right],
\end{equation}

\noindent
where $D^2(a)=a^4_{fct}$. Integrating the last equation, we obtain
\begin{eqnarray}
 &\sigma^{2}(M)=\frac{4\pi\, D^2(a) A_{s}}{k_{0}^{n_s}}\left[ \frac{k^{n_{s}+3}}{n_{s}+3}\Bigg |_{0}^{k_{piv}}+k_{piv}^{n_{s}-n_{b}}\frac{k^{n_{b}+3}}{n_{b}+3}\Bigg |_{k_{piv}}^{k_{R}}\right],\nonumber\\
 &=\frac{4\pi\, a^4_{fct}\,A_{s}}{k_{0}^{n_{s}}}\left[\frac{k_{piv}^{n_{s}+3}}{n_{s}+3}+ \,k_{piv}^{n_{s}-n_{b}}\left(\frac{k_{R}^{n_{b} +3}}{n_{b}+3}-\frac{k_{piv}^{n_{b}+3}}{n_{b}+3}\right) \right],\nonumber\\
 &=\frac{4\pi\, a^4_{fct}\,A_{s}}{k_{0}^{n_{s}}}\, k_{piv}^{n_{s}-n_{b}}\left[\frac{(n_{b}-n_{s})k_{piv}^{n_{b}+3}+(n_{s}+3)k_{R}^{n_{b}+3}}{(n_{s}+3)(n_{b}+3)}\right].
 \label{eq:Ap_sigma_FCT_brk}
\end{eqnarray}

We substitute $k_R$ as given by Eq. \eqref{eq:k_R__fct}, then Eq. \eqref{eq:Ap_sigma_FCT_brk} results in

\begin{equation}
\sigma^{2}(M)=A_{piv}\JS{f_m^{\JSH{\frac{n_b+3}{3}}}}\left(S_{1}\JS{f_m^{-\alpha}}+S_{2} M^{-\frac{(n_{b}+3)}{3}}\right),
\end{equation}
where we included the definitions of $A_{piv}$, $S_{1}$ and $S_{2}$ given by Eqs. \eqref{eq:Apiv_fct} and \eqref{eq:S1 and S2 FCT}, respectively. Furthermore, $\nu(M)$ (Eq. \ref{eq:nu}) in this scenario reads as
\begin{equation}
\nu(M)=\frac{\delta_{c}}{\left[A_{piv}f_m^{\JSH{\frac{n_b+3}{3}}}\left(S_{1}\JS{f_m^{-\alpha}}+S_{2} M^{-\frac{(n_{b}+3)}{3}}\right)\right]^{1/2}}. 
\label{eq: AP nu FCT brk v0}
\end{equation}

By defining a characteristic mass $M_{*}$ as the mass which satisfies $\nu(M_{*})=1$, we have

\begin{equation}
    1 = \frac{\delta_{c}}{\left[A_{piv}f_m^{\JSH{\frac{n_b+3}{3}}}\left(S_{1}\JS{f_m^{-\alpha}}+S_{2} M_{*}^{-\frac{(n_{b}+3)}{3}}\right)\right]^{1/2}},
\end{equation}
\noindent
where $M_{*}$ is then computed as
\begin{equation}
    M_{*}\equiv \left(\frac{\delta_{c}^{2}}{A_{piv}f_m^{\JSH{\frac{n_b+3}{3}}}}\,S_{2}-\frac{S_{1}\JS{f_m^{-\alpha}}}{S_{2}}\right)^{-\frac{3}{n_{b}+3}},
    \label{eq: AP Ms FCT brk}
\end{equation}

\noindent
and $\delta_{c}$ is given by

\begin{equation}
    \delta_{c} = \left[A_{piv}f_m^{\JSH{\frac{n_b+3}{3}}}\left(S_{1}\JS{f_m^{-\alpha}}+S_{2} M_{*}^{-\frac{(n_{b}+3)}{3}}\right)\right]^{1/2}.
    \label{eq:Apdelta_c FCT brk}
\end{equation}

With the definitions in Eqs. \eqref{eq: AP Ms FCT brk} and \eqref{eq:Apdelta_c FCT brk} we rewrite \eqref{eq: AP nu FCT brk v0} as

\begin{equation}
\nu(M)=\left(\frac{S_{1}\JS{f_m^{-\alpha}}+S_{2} M_{*}^{-\frac{(n_{b}+3)}{3}}}{S_{1}\JS{f_m^{-\alpha}}+S_{2} M^{-\frac{(n_{b}+3)}{3}}}\right)^{1/2}, 
\label{eq:AP nu_m}
\end{equation}
and its derivative with respect to $M$, results in

\begin{equation}
\frac{d\nu}{dM}=\frac{S_{2}(n_{b}+3)}{6M^{\frac{(n_{b}+6)}{3}}} \frac{\left(S_{1}\JS{f_m^{-\alpha}}+S_{2} M_{*}^{-\frac{(n_{b}+3)}{3}}\right)^{1/2}}{\left(S_{1}\JS{f_m^{-\alpha}}+S_{2} M^{-\frac{(n_{b}+3)}{3}}\right)^{3/2}}.
\label{eq: AP dnudM FCT brk}
\end{equation}

Replacing Eqs. \eqref{eq:AP nu_m} and \eqref{eq: AP dnudM FCT brk} in Eqs. \eqref{eq: fnu Gaussian} and \eqref{eq:dndm_gen}, we obtain the PBH mass function for this scenario

\begin{eqnarray}
\left(\frac{dn}{dM}\right)_{\text{fct}}^{\text{brk}}&=&\frac{(n_{b}+3)\,S_{2}}{3\,M^{\frac{(n_{b}+9)}{3}}}\frac{\rho_{DM(a)}}{\sqrt{2 \pi}} \frac{\left(S_{1}\JS{f_m^{-\alpha}}+S_{2} M_{*}^{-\frac{(n_{b}+3)}{3}}\right)^{1/2}}{\left(S_{1}\JS{f_m^{-\alpha}}+S_{2} M_{}^{-\frac{(n_{b}+3)}{3}}\right)^{3/2}}\nonumber  \\
&&\times \exp \left[-\frac{1}{2}\frac{S_{1}\JS{f_m^{-\alpha}}+S_{2} M_{*}^{-\frac{(n_{b}+3)}{3}}}{S_{1}\JS{f_m^{-\alpha}}+S_{2} M^{-\frac{(n_{b}+3)}{3}}} \right].
\label{eq: AP dndm FCT brk}
\end{eqnarray}

\subsection{Horizon crossing}\label{sec:appendix HC}

\subsubsection{Standard Power Spectrum}

In this scenario, $\sigma(k)$ is given by Eq. \eqref{eq:sigma_plaw_fct}, were $D^2(a)=a^4_{hc}$ and $a_{hc}$ is given by Eq. \eqref{eq:a_hc}. Replacing $k_R$ by the definition given in Eq. \eqref{eq:kR_hc}, we have

\begin{eqnarray}
    \sigma^2(M) = \frac{4\pi\,A_s}{k_0^{n_{s}}} \left(\frac{2GH_0\sqrt{\Omega_{r,0}}}{c^3}\right)^2\frac{C^{n_s+3}_{hc}}{n_s+3}\JS{\left(\frac{f_m}{M}\right)^{\frac{n_s-1}{2}}}\nonumber\\
    = \frac{4\pi\,A_s}{k_0^{n_{s}}}\left(\frac{G}{\pi\,c^2}\right)^4\frac{C^{n_s+7}_{hc}}{n_s+3}\JS{\left(\frac{f_m}{M}\right)^{\frac{n_s-1}{2}}}.
    \label{eq:AP SigmaM_HC_std}
\end{eqnarray}

With Eq. \eqref{eq:AP SigmaM_HC_std}, $\nu(M)$ is expressed as

\begin{equation}
    \nu(M)=\delta_c\left(\frac{k_0^{n_s}\,(n_s+3)}{4\pi\, A_s\, C^{n_s+7}_{hc}}\right)^{1/2} \left(\frac{\pi\,c^2}{G}\right)^2\,\JS{\left(\frac{M}{f_m}\right)^{\frac{n_s-1}{4}}},
    \label{eq: AP nuM HC std}
\end{equation}

\noindent
where $M_*$ is obtained by defining $\nu(M_*)=1$ and solving for $M_*$. This results in

\begin{eqnarray}
    1=\delta_c\left(\frac{k_0^{n_s}\,(n_s+3)}{4\pi\, A_s\, C^{n_s+7}_{hc}}\right)^{1/2} \left(\frac{\pi\,c^2}{G}\right)^2\,\JS{\left(\frac{M_*}{f_m}\right)^{\frac{n_s-1}{4}}},
\end{eqnarray}
then

\begin{equation}
    M_* = \JS{\frac{f_m}{\delta_c^{\frac{4}{n_s-1}}}}\left[ \sqrt{\frac{4\,\pi\,(A_{s}/k_0^{n_s})}{n_s+3}} \left(\frac{G}{\pi\, c^2}\right)^2 C_{hc}^{\frac{n_s+7}{2}}\right]^\frac{4}{n_s-1}.
    \label{eq: AP Mstar HC std}
\end{equation}
    
We can solve $\delta_c$ as a function of $M_*$ from Eq. \eqref{eq: AP Mstar HC std} as

\begin{equation}
    \delta_c = \left(\frac{4\pi\, A_s\, C^{n_s+7}_{hc}}{k_0^{n_s}\,(n_s+3)}\right)^{1/2} \left(\frac{G}{\pi\,c^2}\right)^2\,\JS{\left(\frac{M_*}{f_m}\right)^{\frac{1-n_s}{4}}},
\end{equation}
and replacing this result in Eq. \eqref{eq: AP nuM HC std} we can rewrite $\nu(M)$ as

\begin{equation}
    \nu(M) = \left(\frac{M}{M*}\right)^{\frac{n_s-1}{4}}.
    \label{eq: AP nuMstar HC std}
\end{equation}

Then, the derivative of $\nu(M)$ with respect to $M$ is given by

\begin{equation}
    \frac{d\nu}{dM} = \frac{n_s-1}{4}\frac{1}{M_*}\left(\frac{M}{M_*}\right)^{\frac{n_s-5}{4}}.
    \label{eq: AP dnudM HC std}
\end{equation}

By replacing these results in Eqs. \eqref{eq: fnu Gaussian} and \eqref{eq:dndm_gen} we obtain

\begin{eqnarray}
    \left(\frac{dn}{dM}\right)_{\text{hc}}^{\text{std}} &=& \frac{\rho_{DM}(a)}{\sqrt{2\pi}}\frac{(n_s-1)}{2\,M^2}\left(\frac{M_*}{M}\right)^{\frac{1-n_s}{4}}  \nonumber\\
    &&\times\exp\left({-\frac{1}{2}\left(\frac{M_*}{M}\right)^{\frac{1-n_s}{2}}}\right).
    \label{eq: AP dndm HC std}
\end{eqnarray}

\subsubsection{Broken Power Spectrum}

Here, $\sigma(k)$ is given by Eq. \eqref{eq:Ap_sigma_FCT_brk}. We follow the same procedure as before, obtaining

\begin{equation}
    \sigma_{hc}^2(M) = A^{'}_{piv}\JS{f_m^{\frac{n_b-1}{2}}}\left[S^{'}_{1}\JS{f_m^{-\alpha'}} \, M^2+ S^{'}_{2}\, M^{\frac{(1-n_{b})}{2}}\right],
    \label{eq: AP sigma HC brk}
\end{equation}

\noindent
where we used $A^{'}_{piv}$, $S^{'}_{1}$ and $S^{'}_{2}$ given by Eqs. \eqref{eq: Apiv_hc} and \eqref{eq: S1 and S2 HC} respectively. $\nu(M)$ is then given by

\begin{equation}
    \nu(M) = \frac{\delta_c}{\sqrt{A^{'}_{piv}\JS{f_m^{\frac{n_b-1}{2}}}}\left[S^{'}_1 \,\JS{f_m^{-\alpha'}} M^2+ S^{'}_2\, M^{\frac{(1-n_{b})}{2}}\right]^{1/2}}.
    \label{eq: AP nuM HC brk}
\end{equation}

With this, we define $M_*$ (satisfying $\nu(M_*)=1$) trough 

\begin{equation}
    \delta_c^2 = A^{'}_{piv}\JS{f_m^{\frac{n_b-1}{2}}}\left[S^{'}_1 \,\JS{f_m^{-\alpha'}}\, M_*^2+ S^{'}_2\, M_*^{\frac{(1-n_{b})}{2}}\right],
    \label{eq: AP Mstar HC brk}
\end{equation}
which must be solved numerically. Then, we write Eq. \eqref{eq: AP nuM HC brk} in terms of $M_*$ as

\begin{equation}
        \nu(M) = \left(\frac{S^{'}_1 \,\JS{f_m^{-\alpha'}}\, M_{*}^2+ S^{'}_2\, M_{*}^{\frac{(1-n_{b})}{2}}}{S^{'}_1 \,\JS{f_m^{-\alpha'}} \, M^2+ S^{'}_2\, M^{\frac{(1-n_{b})}{2}}}\right)^{1/2},
    \label{eq: AP nuMstar HC brk}
\end{equation}
and its derivative with respect to $M$ reads

\begin{eqnarray}
    \frac{d\nu}{dM} = \frac{\left[\left(\frac{n_b-1}{2}\right)S^{'}_2\,M^{\frac{1-n_b}{2}}-2S^{'}_1\,\JS{f_m^{-\alpha'}}\,M\right]}{2}\nonumber\\
    \times\frac{\left(S^{'}_1\,\JS{f_m^{-\alpha'}} \, M_{*}^2+ S^{'}_2\, M_{*}^{\frac{(1-n_{b})}{2}}\right)^{1/2}}{\left(S^{'}_1\,\JS{f_m^{-\alpha'}} \, M^2+ S^{'}_2\, M^{\frac{(1-n_{b})}{2}}\right)^{3/2}}.
    \label{eq: AP dnudM HC brk}
\end{eqnarray}

Using these results, we obtain

\begin{eqnarray}
    \left(\frac{dn}{dM}\right)_{\text{hc}}^{\text{brk}} = \frac{ \rho_{DM}(a)}{\sqrt{2\pi}}\frac{\left[\left(\frac{n_b-1}{2}\right)S^{'}_2\,M^{\frac{1-n_b}{2}}-2S^{'}_1\,\JS{f_m^{-\alpha'}}\,M\right]}{2}\nonumber\\
    \times\frac{\left(S^{'}_1\,\JS{f_m^{-\alpha'}} \, M_{*}^2+ S^{'}_2\, M_{*}^{\frac{(1-n_{b})}{2}}\right)^{1/2}}{\left(S^{'}_1\,\JS{f_m^{-\alpha'}} \, M^2+ S^{'}_2\, M^{\frac{(1-n_{b})}{2}}\right)^{3/2}}\nonumber\\
    \times\exp{\left[-\frac{1}{2}\frac{\left(S^{'}_1\,\JS{f_m^{-\alpha'}} \, M_{*}^2+ S^{'}_2\, M_{*}^{\frac{(1-n_{b})}{2}}\right)}{\left(S^{'}_1\,\JS{f_m^{-\alpha'}} \, M^2+ S^{'}_2\, M^{\frac{(1-n_{b})}{2}}\right)}\right]}.
    \label{eq: AP dndm HC brk}
\end{eqnarray}

\bsp	
\label{lastpage}
\end{document}